\documentclass{cpbtex}
\usepackage{tabularx} 
\usepackage{color} 
\usepackage{epsfig,graphicx,dcolumn,bm,times}
\usepackage{widetext}
\usepackage{amsmath}

\begin{document}

\title{Contagion dynamics on adaptive multiplex networks with awareness-dependent rewiring\thanks{Project supported by the National Natural Science Foundation of China (Grant Nos.~11601294 and 61873154), Shanxi Scholarship Council of China (Grant No. 2016-011), Shanxi Province Science Foundation for Youths (Grant Nos. 201601D021012, 201801D221011, 201901D211159, 201801D221007, 201801D221003), and the 1331 Engineering Project of Shanxi Province.}}


\author{Xiao-Long Peng$^{1,2}$\thanks{Corresponding author. E-mail:~xlpeng@sxu.edu.cn},  ~ Yi-Dan Zhang$^{1,2,3}$\\
$^{1}${Complex Systems Research Center, Shanxi University, Taiyuan 030006, Shanxi, China}\\  
$^{2}${Shanxi Key Laboratory of Mathematical Techniques and Big Data Analysis on Disease}\\
{Control and Prevention, Shanxi University, Taiyuan 030006, Shanxi, China}\\ 
$^{3}${School of Mathematical Sciences, Shanxi University, Taiyuan 030006, Shanxi, China}\\
}   


\date{\today}
\maketitle

\begin{abstract}
Over the last few years, the interplay between contagion dynamics of social influences (e.g., human awareness, risk perception, and information dissemination) and biological infections has been extensively investigated within the framework of multiplex networks. The vast majority of existing multiplex network spreading models typically resort to heterogeneous mean-field approximation and microscopic Markov chain approaches. Such approaches usually manifest richer dynamical properties on multiplex networks than those on simplex networks; however, they fall short of a subtle analysis of the variations in connections between nodes of the network and fail to account for the adaptive behavioral changes among individuals in response to epidemic outbreaks. To transcend these limitations, in this paper we develop a highly integrated effective degree approach to modeling epidemic and awareness spreading processes on multiplex networks coupled with awareness-dependent adaptive rewiring. This approach keeps track of the number of nearest neighbors in each state of an individual; consequently, it allows for the integration of changes in local contacts into the multiplex network model. We derive a formula for the threshold condition of contagion outbreak; also, we provide a lower bound for the threshold parameter to indicate the effect of adaptive rewiring. The threshold analysis is confirmed by extensive simulations. Our results show that awareness-dependent link rewiring plays an important role in enhancing the transmission threshold as well as lowering the epidemic prevalence. Moreover, it is revealed that intensified awareness diffusion in conjunction with enhanced link rewiring makes a greater contribution to disease prevention and control. In addition, the critical phenomenon is observed in the dependence of the epidemic threshold on the awareness diffusion rate, supporting the metacritical point previously reported in the literature. This work might shed light on the understanding of the interplay between epidemic dynamics and social contagion on adaptive networks.
\end{abstract}



\section{Introduction}
Nowadays, the communication networks---through media, such as telephones, emails, Facebook, Twitter and Wechat---provide people easy access to information about the emergence or outbreaks of infectious diseases. In many cases, public health communications concerning the risk of infections have a beneficial effect on human behavior \cite{Bauc13}. For example, during the COVID-19 pandemic, the World Health Organization has disseminated information on health issues such as quarantining, social distancing, regulating international travelers, closing schools, and operating public transportation and workplaces. People who receive the disease information become aware of the transmission risk and tend to change their behavior to reduce their susceptibility to infections \cite{Furg07}. In this way, social contagion \cite{Ruan15,Wangwei15, Xu16} of human awareness about infections influences the biological contagion \cite{Bauc13}.

In the literature, many works have been devoted to understanding the impact of human awareness on infectious disease spread. For example, Funk \textit{et al.} \cite{Funk09} formulated a network model to study how the spread of awareness, prompted by first-hand information about a disease case, affects the disease spread. Along the same lines, Wu \textit{et al.} \cite{Wu12} proposed a continuous heterogeneous mean-field model to study the impact of contact, local and global awareness on the epidemic spread on scale-free networks. These models are based on a simplex network structure by simply assuming that the physical contact network layer (for disease transmission) and the information network layer (for awareness propagation) are identical in that their connections among the individuals coincide with each layer. In reality, however, the information network, where the individuals communicate information about disease cases, is often topologically different from the face-to-face contact network, where the disease can be transmitted through infectious contacts. To address this problem, Granell \textit{et al.} \cite{Gran13} pioneered the study of competing dynamics between awareness and disease spreading processes on a multiplex network consisting of two network layers, in which the nodes represent the same entities in both layers while edges in different layers stand for different types of social relationships \cite{Gomez15}. Their multiplex network framework was applied to a susceptible-infected-susceptible (SIS) \cite{AM91} disease transmission that confers no immunity and a unaware-aware-unaware (UAU) information diffusion process. Granell \textit{et al.} \cite{Gran13} utilized a microscopic Markov chain approach to obtain the threshold condition for the disease outbreak; interestingly, they discovered a metacritical point for the epidemic threshold. They further considered the general case where the awareness does not guarantee complete immunization and infection does not confer instant awareness of it; also, the impact of mass media on the epidemic prevalence was incorporated \cite{Gran14}.

The pioneering work by Granell \textit{et al.} has prompted a number of supplementary studies on the competing processes of disease and awareness on multiplex networks. For instance, Guo \textit{et al.} \cite{Guo15} proposed a local awareness-controlled SIS epidemic model on multiplex networks; interestingly, they observed two-stage effects on epidemic threshold and epidemic prevalence. Kan and Zhang \cite{Zhang17} investigated the combined effects of awareness spread and self-initiated awareness behavior on the SIS disease dynamics. More recently, Xia \textit{et al.} \cite{Xia19,WXia19,Xia18} extended the multiplex network model of awareness-disease dynamics to a susceptible-infected-recovered (SIR) \cite{AM91} epidemic process that allows permanent immunity after infection. All of these models have employed the microscopic Markov chain approach by Granell \textit{et al.} \cite{Gran13,Gran14} to identify the epidemic threshold as the largest eigenvalue of a matrix operator that includes information about the network structure and the density of aware individuals (who know about the infection risk) in the steady state \cite{Guo15,Zhang17,Xia19,WXia19,Xia18}.

The above models considered the impact of behavioral response to disease information by simply integrating the impact into a reduced susceptibility for aware susceptibles (i.e., susceptible individuals who are aware of the disease). Nevertheless, they did not actually account for the human behavioral changes that can reshape the underlying contact network, which in turn results in the coevolution of epidemics and network topology \cite{Gross06,Gross08,Shaw08,Marceau10,Segb10,Kamp10,WangBin11,PAM12,Juher13,Peng16,Kiss18,Zhangxg19,Lu19}. The failure of the heterogeneous mean-field approximation and microscopic Markov chain approaches to reflect the network evolution (due to behavioral changes in response to infection threat) is rooted in the lack of time evolution of edges in the total-degree-based approximation frameworks. To eliminate this shortcoming, Gross \textit{et al.} \cite{Gross06} proposed an adaptive network epidemic model with adaptive rewiring of dangerous links, in which susceptible nodes are allowed to break away from their infected neighbors and reconnect towards other noninfected nodes. Despite the low complexity in the analytical formalism, this adaptive epidemic model revealed rich dynamical properties, such as bistability, oscillations and first-order phase transitions \cite{Gross06,Zhangxg19}. However, the low-dimensional formalism, in general, does not provide with high accuracy the time evolution of the disease progression and the underlying network topology. To this end, Marceau \textit{et al.} \cite{Marceau10} introduced an improved compartmental formalism in which nodes are classified into groups according to their total degree (namely, the total number of their neighbors) and their infectious degree (namely, the number of their infected neighbors). Since this formalism features the number of active neighbors associated with the state of infectiousness, in the rest of this paper we shall call it the effective degree formalism. Cai \textit{et al.} \cite{Cai16} successfully addressed the dynamical correlation problem of SIS epidemic model on complex networks by combing the heterogeneous mean-field theory \cite{Pastor15,FuSmall14,WangWei17} and the effective degree model. Ma \textit{et al.} \cite{Ma11} applied the effective degree formalism to both the SIS and the SIR epidemic models and found that these two effective degree epidemic models give rise to different epidemic thresholds. Zhou \textit{et al.} \cite{Zhou19} has recently utilized the effective degree formalism in the study of disease and awareness spreading dynamics on multiplex networks, where excellent agreement was observed between theoretical predictions and stochastic simulations for the spreading dynamics. The authors further developed a partial effective degree theory with a mixture of the effective degree formalism for the contact network layer and the heterogeneous mean-field approximation for the information network layer \cite{Zhou19}. In their work, however, the behavioral adaptations of individuals that could lead to network evolution have not been incorporated in the effective degree spreading model on multiplex networks.

In view of the practical consideration that an individual's behavioral adaptations (e.g., activity reduction \cite{Kotnis13,Rizzo14}) could be affected by their own infection state and their awareness of disease information \cite{Kotnis13,Ding18}, it is of significance to integrate the awareness-related behavioral adaptations into the spreading dynamics of epidemic and awareness on multiplex networks. In this paper, we apply the effective degree approach to modeling the simultaneous contagion dynamics of an SIS epidemic transmission and a UAU awareness propagation on adaptive multiplex networks with awareness-dependent rewiring in the physical contact network layer. Our model is described in section \textcolor{blue}{2} and formalised in section \textcolor{blue}{3}. Threshold conditions of disease outbreak and awareness information invasion are analyzed for the full effective degree model in section \textcolor{blue}{4} and the lower bounds of the threshold conditions are derived for a reduced model with some special case in section \textcolor{blue}{5}. Simulation results are presented in section \textcolor{blue}{6}, where the effects of adaptive rewiring and awareness diffusion on the epidemic threshold and epidemic prevalence are analyzed and the critical phenomenon in the reliance of epidemic onset on awareness diffusion is highlighted. Our conclusions are drawn in section \textcolor{blue}{7}.

\section{The SIS-UAU contagion model on adaptive multiplex networks}\label{sec2}

\begin{figure}
  \includegraphics[width=\columnwidth]{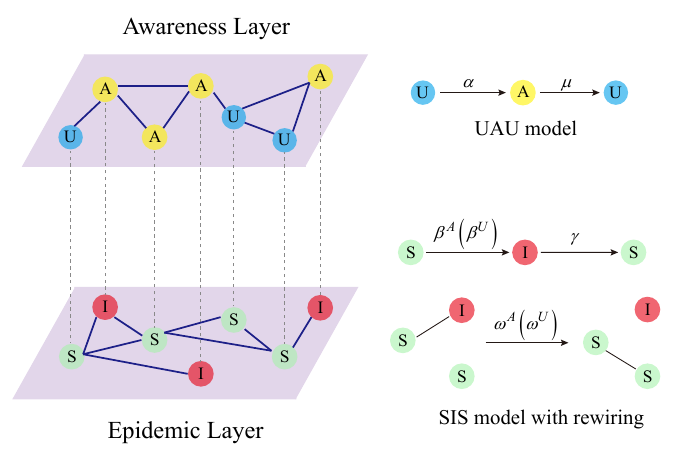}
  \caption{Flowchart of the multiplex network spreading model of an SIS epidemic process and a UAU awareness propagation process, where the events of adaptive rewiring of dangerous (infectious) contacts (that connect an infected node and a susceptible one) randomly occurs in the epidemic layer. In the awareness layer, unaware nodes become aware after they are informed of disease information by an aware neighbor at a rate $\alpha$, whereas an aware node returns to being unaware again after it forgets the disease information at a rate $\mu$. In the epidemic layer, susceptible nodes are infected by their infected neighbors at a transmission rate that depends on the awareness state of the susceptible; here we denote by $\beta^{A}$ (resp., $\beta^{U}$) the transmission rate for aware (resp., unaware) susceptible nodes. Meanwhile, each infected node recovers to become susceptible again at a rate $\gamma$. In addition, for each infectious connection linking an infected node and a susceptible one, the susceptible will break the connection away from the infected and rewire towards a randomly selected susceptible (other than the original susceptible and its existing neighbors) at a rewiring rate, which is dependent on the awareness state of the original susceptible node. Here we denote by $\omega^{A}$ (resp., $\omega^{U}$) the rewiring rate for the aware (resp., unaware) susceptible nodes.}\label{fig1}
\end{figure}
We start with a two-layer multiplex network, as shown in Fig.~\ref{fig1}, consisting of an awareness (information) layer and an epidemic (physical contact) layer.  The awareness layer represents the online social (or communication) network where individuals propagate disease information to their neighbors on the same layer, while the epidemic layer stands for the physical contact network on which individuals transmit disease when infected individuals have a physical contact with susceptible ones. The nodes between the two layers form a one-to-one correspondence as illustrated by the dashed lines in Fig.~\ref{fig1}; each dashed line indicates a correspondence for the same node. Namely, both layers contain exactly the same nodes. However, the connectivity patterns in the two layers are different from each other since the edges in different layers represent different types of relationships. That means the degree distributions are distinct between the two layers.

In the awareness layer we adopt the UAU model proposed by Granell \textit{et al.} \cite{Gran13} for the diffusion of awareness information on the disease, where each individual is in one of the two discrete states: aware (A) or unaware (U), representing whether or not they have acquired the disease information. Unaware individuals become aware at a diffusion rate $\alpha$ by communication with their aware neighbors (where the notations in the model definition are summarized in Table~\ref{table1}). In addition, unaware individuals will directly become aware with probability $1$ once they get infected via disease-causing physical contacts. Meanwhile, each aware individual becomes unaware again at a rate $\mu$ after it forgets the awareness or no longer cares about it.

\begin{table*}[htbp]
 \addtolength{\tabcolsep}{0.5pt}
 \caption{Notation used in the model definition. }\label{table1}
 \hspace{-0.15cm}
 \begin{tabularx}{1\textwidth}{lX}
 \hline
 Term & Meaning\\
 \hline
 $\alpha$ & awareness information diffusion rate\\
 $\mu$ & awareness information forgetting rate\\
 $\beta^U$ & disease transmission rate for unaware susceptible individuals\\
 $\beta^A$ & disease transmission rate for aware susceptible individuals\\
 $\beta$ & disease transmission rate (note that in this paper we simply assume a common transmission rate, i.e., $\beta^U=\beta^A=\beta$) \\
  $\gamma$ & recovery rate from infection\\
 $\omega^U$ & link rewiring rate on each risky contact between a unaware susceptible individual and an infected one\\
  $\omega^A$ & link rewiring rate on each risky contact between a aware susceptible individual and an infected one\\
  $y$ & ratio between $\omega^U$ and $\omega^A$, i.e., $y=\omega^U/\omega^A$ \\
 \hline
 \end{tabularx}
\end{table*}

In the epidemic layer we consider an SIS epidemic process with awareness-dependent link rewiring, where each individual is in one the the two distinct sates: susceptible (S) or infected (I). The transmission rate for aware (resp., unaware) susceptible nodes is $\beta^{A}$ (resp., $\beta^{U}$), the recovery rate for each infected node to return to the susceptible sate is $\gamma$, and the adaptive rewiring rate for the aware (resp., unaware) susceptible node belonging to each susceptible-infected ($S$-$I$) edge to rewire away from the infected node and connect towards another randomly selected susceptible node is $\omega^{A} (\omega^{U})$. Since awareness of disease infection often induces individuals to take preventive measures to evade or reduce the risk of infection in the epidemic layer \cite{Paolo14}, it is natural to assume $\beta^{A}\leq\beta^{U}$ and $\omega^{A}\geq \omega^{U}$. For the sake of simplicity, we set $\omega^{U}=y\omega^{A}$ with $0\leq y\leq1$; here $y=0$ means unaware individuals take no self-protective measures at all, whereas $y=1$ means the rewiring rate for susceptible nodes is independent of their awareness state. It is worth mentioning that in the present paper we simply assume $\beta^A=\beta^{U}=\beta$ and $\omega^A>\omega^U$. That is, we don't differentiate the transmission rate between aware and unaware susceptibles, since the impact of awareness information on reducing the risk of infection has already been reflected in the larger rewiring rate for aware susceptibles than unaware ones. Moreover, in our model the link rewiring process and the disease transmission process take place simultaneously but independently.

Note that most existing concurrent spreading models on multiplex networks have assumed a fixed network structure in each layer, whereas in our model the network structure in the epidemic layer coevolves with the epidemic dynamics due to adaptive rewiring of disease-causing contacts. In this sense we refer to our model as the SIS-UAU contagion model on adaptive multiplex networks. For the sake of simplicity, in the present work we only consider the network evolution of the epidemic layer and hypothesize that the network topology in the awareness layer does not change during the process of the concurrent spreading of epidemic and awareness, though in reality the connectivity pattern in the awareness layer may also evolve with time.

\begin{table*}[htbp]
 \addtolength{\tabcolsep}{0.5pt}
 \caption{Notation used in the model formulation and analytical approximation. }\label{table2}
 \hspace{-0.15cm}
 \begin{tabularx}{1\textwidth}{lX}
 \hline
 Term & Meaning\\
 \hline
 $N$ & number of nodes in each layer (note that each node in one layer forms a replica in the other layer) \\
 $k$ & total degree (i.e., the number of neighbors) of a node in the epidemic layer. $k=1,2,\cdots, K$, where $K$ is the maximal degree\\
 $i$ & infected degree (i.e., the number of infected neighbors) of a node in the epidemic layer, $0\leq i\leq k$\\
 $n$ & total degree (i.e., the number of neighbors) of a node in the awareness layer. $n=1,2,\cdots, M$, where $M$ is the maximal degree\\
 $a$ & aware degree (i.e., the number of aware neighbors) of a node in the epidemic layer, $0\leq a\leq n$\\
 $S_{ki}^{U_{na}}$ & fraction of unaware susceptible nodes both with $k$ total neighbors and $i$ infected neighbors in the epidemic layer but also with $n$ total neighbors and $a$ aware neighbors in the awareness layer\\
 $S_{ki}^{A_{na}}$ & fraction of aware susceptible nodes both with $k$ total neighbors and $i$ infected neighbors in the epidemic layer but also with $n$ total neighbors and $a$ aware neighbors in the awareness layer\\
 $I_{ki}^{A_{na}}$ & fraction of infected nodes both with $k$ total neighbors and $i$ infected neighbors in the epidemic layer but also with $n$ total neighbors and $a$ aware neighbors in the awareness layer\\
 $S^U$ & fraction of unaware susceptible nodes  \\
 $S^A$ & fraction of aware susceptible nodes \\
 $I^A$ & fraction of infected nodes (that are automatically aware about the disease due to infection)\\
 $\rho^X$ & density of nodes in state $X$, where $X\in\{S,I,U,A\}$, note that $\rho^S=S^A+S^U=S$ \\
 $X_Y$ & fraction of $X$-$Y$ links emanating from nodes in $X$ state towards nodes in $Y$ state, where $X,Y\in\{S,I,U,A\}$ \\
 $X_{YZ}$ & fraction of triplets of type $Y$-$X$-$Z$ where a node in $Y$ is connected to a node in $X$ state, which in turn is connected to another node in $Z$ state, where $X,Y,Z\in\{S,I,U,A\}$ \\
 \hline
 \end{tabularx}
\end{table*}

\section{Effective degree model formulation}\label{sec3}

According to the evolution mechanism of the dynamical processes described in the previous section, all the nodes in the multiplex network can be grouped into three different states: susceptible and unaware (SU), susceptible and aware (SA), or infected and aware (IA). Following the effective degree approach by Marceau \textit{et al.} \cite{Marceau10} and Lindquist \textit{et al.} \cite{Ma11}, every node in the multiplex network is classified by its disease state, awareness state, and its \textit{total degree} in each layer, as well as by its \textit{infected degree} (in the epidemic layer) and its \textit{aware degree} (in the awareness layer). Here, by total degree in each layer we mean the total number of links connected to a node in the layer and by infected degree and aware degree we mean the number of links shared with infected individuals and those shared with aware individuals, respectively. Therefore, each node can be categorized into classes (also called compartments in the literature) of ${\text{X}}_{ki}{{\text{Y}}_{na}}$, where $\text{X}\in \{\text{S}, \text{I}\}$ and $\text{Y}\in \{\text{U}, \text{A}\}$, subscripts $k$ and $i$ denote the node's total degree and infected degree, respectively, in the epidemic layer, and subscripts $n$ and $a$ denote the node's total degree and aware degree, respectively, in the awareness layer. Obviously, one has $i\leq k$ and $a\leq n$ according to the specific definition of the degrees. Note that the fake class of $\text{I}_{ki}{\text{U}_{na}}$ actually does not exist in our model because according to the definition of our model, once an individual gets infected it becomes aware of the disease immediately. We denote by $X_{ki}^{Y_{na}}(t)$ the fraction of nodes in class $\text{X}_{ki}{\text{Y}_{na}}$ at time $t$ (where the notations used in the model formulation are summarized in Table~\ref{table2}). For the sake of readability, hereinafter we drop the explicit time dependence of $S_{ki}^{U_{na}}$, $S_{ki}^{A_{na}}$, $I_{ki}^{A_{na}}$, and associated variables. Obviously, the conservation law gives rise to
\begin{equation}\label{eq1}
\sum_{k=1}^{K}\sum_{i\leq k}\sum_{n=1}^{M}\sum_{a\leq n}(S_{ki}^{U_{na}}+S_{ki}^{A_{na}}+I_{ki}^{A_{na}})=1,
\end{equation}
where $K$ (resp., $M$) represents the maximal degree in the epidemic (resp., awareness) layer. Following the notations by Marceau \textit{et al.} \cite{Marceau10}, we define the zeroth-order moments of the $S_{ki}^{U_{na}}$, $S_{ki}^{A_{na}}$, $I_{ki}^{A_{na}}$ distributions by
\begin{widetext}
\begin{equation}
S^{U}\equiv \sum_{k,i}\sum_{n,a}S_{ki}^{U_{na}},\quad S^{A}\equiv \sum_{k,i}\sum_{n,a}S_{ki}^{A_{na}},\quad
I^{A}\equiv \sum_{k,i}\sum_{n,a}I_{ki}^{A_{na}},\label{eq2}
\end{equation}
\end{widetext}
the first-order moments by
\begin{widetext}
\begin{eqnarray}\label{eq3}
S_{S}&\equiv &\sum_{k,i}\sum_{n,a}(k-i)\big(S_{ki}^{U_{na}}+S_{ki}^{A_{na}}\big), \quad S_{I} \equiv \sum_{k,i}\sum_{n,a}i \big(S_{ki}^{U_{na}}+S_{ki}^{A_{na}}\big),\nonumber \\
I_{S}&\equiv &\sum_{k,i}\sum_{n,a}(k-i) I_{ki}^{A_{na}}, \quad I_{I}\equiv\sum_{k,i}\sum_{n,a}i I_{ki}^{A_{na}},\nonumber \\
{U}_{U}&\equiv &\sum_{k,i}\sum_{n,a}(n-a) S_{ki}^{U_{na}}, \quad {U}_{A}\equiv\sum_{k,i}\sum_{n,a}a S_{ki}^{U_{na}},\nonumber\\
{A}_{U}&\equiv & \sum_{k,i}\sum_{n,a}(n-a)(S_{ki}^{A_{na}}+I_{ki}^{A_{na}}), \quad {A}_{A}\equiv  \sum_{k,i}\sum_{n,a}a(S_{ki}^{A_{na}}+I_{ki}^{A_{na}}),
\end{eqnarray}
\end{widetext}
and the second-order moments by
\begin{widetext}
\begin{eqnarray}\label{eq4}
S_{SI} &\equiv& \sum_{k,i}\sum_{n,a}i(k-i)\big(S_{ki}^{U_{na}}+S_{ki}^{A_{na}}\big),\quad
S_{II} \equiv  \sum_{k,i}\sum_{n,a}i(i-1)\big(S_{ki}^{U_{na}}+S_{ki}^{A_{na}}\big), \nonumber\\
{U}_{UA}&\equiv & \sum_{k,i}\sum_{n,a}a(n-a)S_{ki}^{U_{na}}, \quad {U}_{AA}\equiv\sum_{k,i}\sum_{n,a}a(a-1)S_{ki}^{U_{na}}.
\end{eqnarray}
\end{widetext}
Here, the physical meaning of the zeroth-order moments in Eq.~(\ref{eq2}) is in essence the density of SU, SA and IA nodes. For convenience, let us denote by $\rho^S$, $\rho^I$, $\rho^U$ and $\rho^A$ the density of susceptible, infected, unaware and aware nodes, respectively, in the network. Then it is straightforward to get the following relations:
\begin{equation}\label{eq5}
\rho^S=S=S^{U}+S^{A},\quad \rho^I=I^{A}, \quad \rho^U=S^U, \quad \rho^A=S^A+I^A.
\end{equation}
The first-order moments in Eq.~(\ref{eq3}) represent the density per node of the various types of arcs. (Here, the arcs are defined as the directional links between a pair of nodes. Therefore each undirected edge in the network actually contains two directional arcs \cite{Marceau10}.) For example, the quantity $S_{S}$ denotes the density of $S$-$S$ links emanating from susceptible nodes, and $U_{A}$ denotes the density of $U$-$A$ links emanating from unaware nodes. The second-order moments in Eq.~(\ref{eq4}) correspond to the density per node of the various types of triplets in the entire network. A triplet is referred to as the union of two directional links surrounding the same central node \cite{Marceau10}. For example, the quantities $S_{SI}$ and $U_{AA}$ denote the density of triplets of types $S$-$S$-$I$ and $A$-$U$-$A$ with the central node in class S and U, respectively. Clearly, it follows from Eqs.~(\ref{eq1}), (\ref{eq2}) and (\ref{eq5}) that the conservation of nodes
\begin{equation}\label{eq6}
\rho^S+\rho^I=1,  \quad \rho^U+\rho^A=1,
\end{equation}
and the conservation of links
\begin{equation}\label{eq7}
S_S+S_I+I_S+I_I=\langle k\rangle, \quad U_U+U_A+A_U+A_A=\langle n\rangle,
\end{equation}
must hold at any time \cite{Gross06,Marceau10}, where $\langle k\rangle$ and $\langle n\rangle$ are the average degree in the epidemic and awareness layer, respectively. Furthermore, since the network under consideration is undirected, the density of $S$-$I$ links must equal the density of $IS$ links; similarly, the density of $UA$ links is equal to the density of $AU$ links. This leads to the following constraints \cite{Marceau10,Ma11}
\begin{equation}\label{eq8}
S_I=I_S,\quad U_A=A_U.
\end{equation}

Based on our model definition stated in section \textcolor{blue}{2}, at each time step, nodes in each class undergo the following three types of variations. First, nodes can change their state and move to the corresponding class with the same subscripts. Second, nodes can move to other classes due to changes of the state of their neighbors in either network layer. Third, nodes can change to other classes due the effects of adaptive rewiring of infectious contacts in the epidemic layer.

\begin{figure}
  \begin{center}
  \includegraphics[width=\columnwidth]{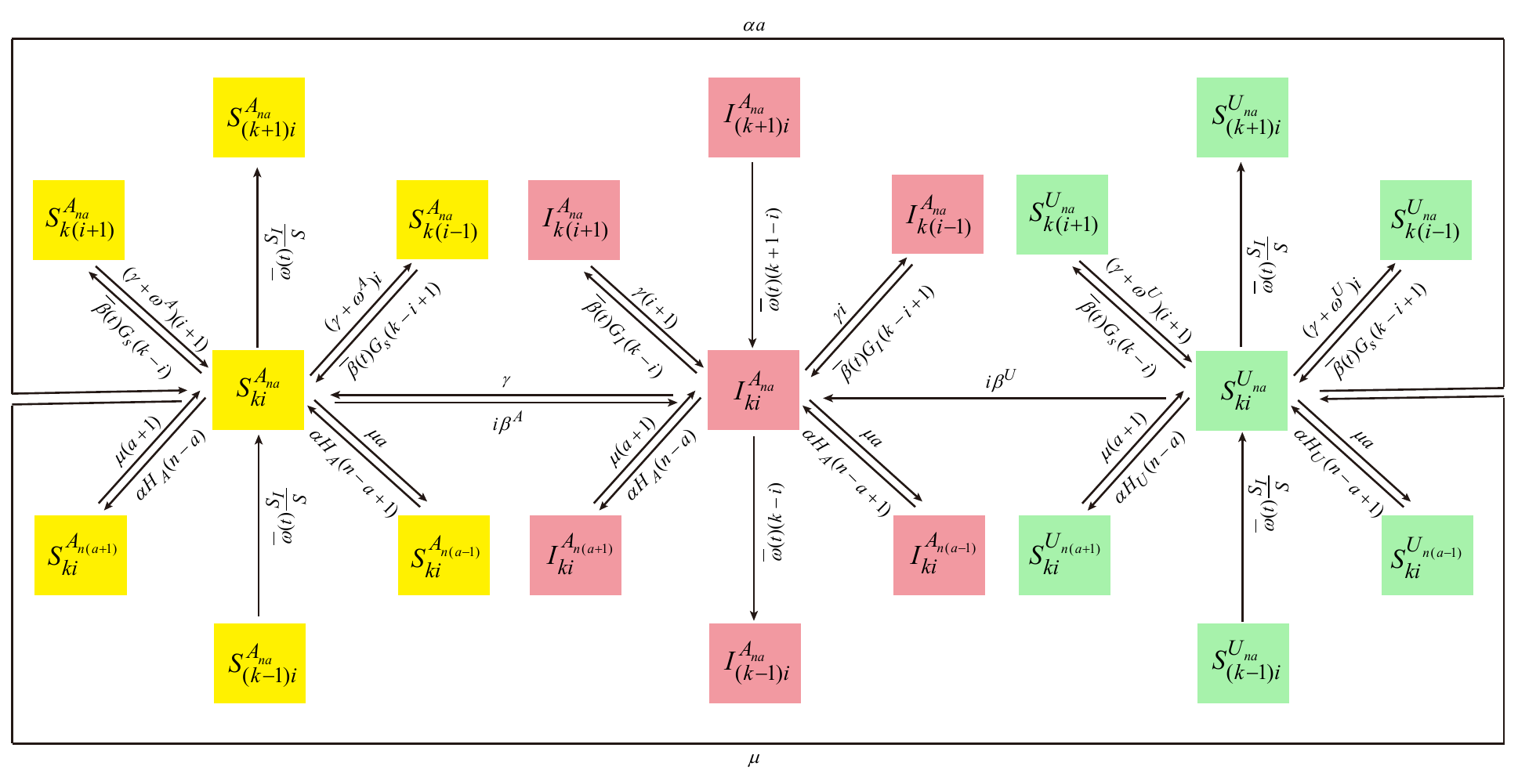}
  \caption{(Color online) Flowchart of the effective degree SIS-UAU contagion model on adaptive multiplex networks. Each arrow denotes the direction of an event that causes a node to move from one compartment to another. The mathematical expression labeled along each arrow is a per node rate for the transition denoted by the arrow. Note that the notations $\bar{\beta}$ and $\bar{\omega}$ are referred to as the average infection rate and average rewiring rate, respectively, for a random susceptible node. Note that, the infection rate of a randomly selected susceptible neighbor of a node in compartment $S_{ki}^{U_{na}}$, depends on its awareness state, which is unknown to us. Thus we approximate it by an average $\bar{\beta}=\beta^A({S^A}/{S})+\beta^U({S^U}/{S})$, where the term $S^A/S$ stands for the probability that the susceptible node is aware and $S^U/S$ for the probability of being unaware. (Note that in this paper we assume $\beta^A=\beta^U=\beta$ since the effects of awareness information on reducing the transmission risk has been included in the larger rewiring rate for aware nodes in comparison to unaware ones, $\omega^A>\omega^U$. Thus we have $\bar{\beta}=\beta$.) Similarly, we have $\bar{\omega}=\omega^A({S^A}/{S})+\omega^U({S^U}/{S})$. Moreover, some notations are simplified by $G_S=S_{SI}/S_{S}$, $G_I=1+S_{II}/S_{I}$, $H_U=U_{UA}/U_{U}$ and $H_A=1+U_{AA}/U_{A}$, which denote the average number of infected (or aware) neighbors one node has given that the node is connected with a node in a certain class.}\label{fig2}
  \end{center}
\end{figure}

According to the flow chart for the SIS-UAU transmission process presented in Fig.~\ref{fig2}, we obtain the set of $\sum_{k=1}^{K}\sum_{n=1}^{M}3(k+1)(n+1)=3KM(K+3)(M+3)/4$ differential equations that govern the temporal evolution of the nodes in each class for $\{(k,i; n,a):1\leq k\leq K,0\leq i\leq k; 1\leq n\leq M,0\leq a\leq n\}$ as follows.
\begin{widetext}
\begin{eqnarray}
\frac{{\rm d} S_{ki}^{U_{na}}}{{\rm d}t}&=&\mu S_{ki}^{A_{na}}-\alpha aS_{ki}^{U_{na}}+\mu\bigg[(a+1)S_{ki}^{U_{n(a+1)}}-aS_{ki}^{U_{na}}\bigg]
 +\alpha \frac{U_{UA}}{U_U}\bigg[(n-a+1)S_{ki}^{U_{n(a-1)}}-(n-a)S_{ki}^{U_{na}}\bigg]\nonumber\\
 &&-i\beta S_{ki}^{U_{na}}+\gamma\bigg[(i+1)S_{k(i+1)}^{U_{na}}-iS_{ki}^{U_{na}}\bigg]
 +\beta\frac{S_{SI}}{S_S}\bigg[(k-i+1)S_{k(i-1)}^{U_{na}}-(k-i)S_{ki}^{U_{na}}\bigg]\nonumber\\
 &&+\omega^{U}\bigg[(i+1)S_{k(i+1)}^{U_{na}}-iS_{ki}^{U_{na}}\bigg]+\bigg(\omega^A\frac{S^A}{S}
 +\omega^U\frac{S^U}{S}\bigg)\frac{S_{I}}{S}\bigg[S_{(k-1)i}^{U_{na}}-S_{ki}^{U_{na}}\bigg], \label{eq9} \\
\frac{{\rm d} S_{ki}^{A_{na}}}{{\rm d}t}&=&-\mu S_{ki}^{A_{na}}+\alpha aS_{ki}^{U_{na}}+\mu\bigg[(a+1)S_{ki}^{A_{n(a+1)}}-aS_{ki}^{A_{na}}\bigg]
 +\alpha \bigg(1+\frac{U_{AA}}{U_A}\bigg)\bigg[(n-a+1)S_{ki}^{A_{n(a-1)}}-(n-a)S_{ki}^{A_{na}}\bigg]\nonumber\\
 &&-i\beta S_{ki}^{A_{na}}+\gamma\*I_{ki}^{A_{na}}+\gamma\bigg[(i+1)S_{k(i+1)}^{A_{na}}-iS_{ki}^{A_{na}}\bigg] +\beta\frac{S_{SI}}{S_S}\bigg[(k-i+1)S_{k(i-1)}^{A_{na}}-(k-i)S_{ki}^{A_{na}}\bigg]\nonumber\\
 &&+\omega^{A}\bigg[(i+1)S_{k(i+1)}^{A_{na}}-iS_{ki}^{A_{na}}\bigg]+\bigg(\omega^A\frac{S^A}{S}
 +\omega^U\frac{S^U}{S}\bigg)\frac{S_{I}}{S}\bigg[S_{(k-1)i}^{A_{na}}-S_{ki}^{A_{na}}\bigg],\label{eq10} \\
\frac{{\rm d} I_{ki}^{A_{na}}}{{\rm d}t}&=&\mu\bigg[(a+1)I_{ki}^{A_{n(a+1)}}-aI_{ki}^{A_{na}}\bigg]+\alpha \bigg(1+\frac{U_{AA}}{U_A}\bigg)\bigg[(n-a+1)I_{ki}^{A_{n(a-1)}}-(n-a)I_{ki}^{A_{na}}\bigg]\nonumber\\
 &&+i\beta\Big(S_{ki}^{U_{na}}+S_{ki}^{A_{na}}\Big)
 +\beta\bigg(1+\frac{S_{II}}{S_I}\bigg)\bigg[(k-i+1)I_{k(i-1)}^{A_{na}}
 -(k-i)I_{ki}^{A_{na}}\bigg]\nonumber\\
 &&-\gamma\*I_{ki}^{A_{na}}+\gamma\bigg[(i+1)I_{k(i+1)}^{A_{na}}-iI_{ki}^{A_{na}}\bigg]+\bigg(\omega^A\frac{S^A}{S}
 +\omega^U\frac{S^U}{S}\bigg)\bigg[(k+1-i)I_{(k+1)i}^{A_{na}}-(k-i)I_{ki}^{A_{na}}\bigg]. \label{eq11}
\end{eqnarray}
\end{widetext}
We call the system of Eqs.~(\ref{eq9})--(\ref{eq11}) the SIS-UAU \textit{effective degree} model. On the right hand side (rhs) of Eq.~(\ref{eq9}), the first term accounts for the addition of $S_{ki}^{U_{na}}$ at rate $\mu S_{ki}^{A_{na}}$ as nodes in compartment $S_{ki}^{A_{na}}$ lose the awareness on disease and become unaware again. The second term indicates that nodes are removed from $S_{ki}^{U_{na}}$ at rate $\alpha aS_{ki}^{U_{na}}$ as they are informed of disease risk by their aware neighbors. The third term accounts for a change in the state of the node's aware neighbors. Nodes from $S_{ki}^{U_{n(a+1)}}$ (resp., $S_{ki}^{U_{na}}$) are transferred to $S_{ki}^{U_{na}}$ (resp., $S_{ki}^{U_{n(a-1)}}$) as one of their aware neighbors becomes unaware again. The fourth term considers the possibility that nodes from $S_{ki}^{U_{n(a-1)}}$ (resp., $S_{ki}^{U_{na}}$) enters the compartment $S_{ki}^{U_{na}}$ (resp., $S_{ki}^{U_{n(a+1)}}$) as one of the node's unaware neighbors becomes aware due to information propagation. Note that we have used here the quantity $U_{UA}/U_{U}$ to measure the average number of aware neighbors that a unaware node reached by following a $U$-$U$ link has, for which we have assumed no degree correlation. The explanation for this is as follows. Since we do not know the exact number of aware nodes that are connected to each of the $(n-a)$ unaware neighbors of nodes in the $S_{ki}^{U_{na}}$ compartment, one alternative is to take the average over the whole network. Under the assumption of degree uncorrelation, a node from $S_{k'i'}^{U_{n'a'}}$ with $a'$ aware neighbors will be reached by following a $U$-$U$ link with probability $(n'-a')S_{k'i'}^{U_{n'a'}}/U_{U}$. Consequently, it follows from Eq.~(\ref{eq4}) that the average number of aware neighbors connected to a unaware node reached by following a $U$-$U$ link is $U_{UA}/U_U$. Thus, the rate associated with the transition from $S_{ki}^{U_{na}}$ to $S_{ki}^{U_{n(a-1)}}$ is $\alpha(U_{UA}/U_U)(n-a)S_{ki}^{U_{na}}$. The fifth term on the rhs of Eq.~(\ref{eq9}) represents the reduction of $S_{ki}^{U_{na}}$ because each node in this compartment is infected at rate $i\beta^U$. The sixth term refers to the change as a result of recovery of one of the node's infected neighbors. The seventh term describes the infection of one of the node's susceptible neighbors. Nodes from $S_{k(i-1)}^{U_{na}}$ (resp., $S_{ki}^{U_{na}}$) are moved to $S_{ki}^{U_{na}}$ (resp., $S_{k(i+1)}^{U_{na}}$) when one of their susceptible neighbors contracts the disease. By analogy with the elucidation for the fourth term, the quantity $S_{SI}/S_S$ gives the average number of infected neighbors of a susceptible node that is reached by following a $S$-$S$ link. Since we do not know the awareness state of each susceptible neighbor of nodes in compartment $S_{ki}^{U_{na}}$, on which the infection rate strongly depends, we considers an average infection rate $\bar{\beta}$ (see Fig.~\ref{fig2}) approximated by an expectation over the entire network. That is, $\bar{\beta}=\beta^A({S^A}/{S})+\beta^U({S^U}/{S})$, where the term $S^A/S$ stands for the probability that the susceptible node is aware and $S^U/S~(=1-S^A/S)$ for the probability of being unaware. The eighth term is attributable to the fact that when one infected neighbor of a susceptible node is rewired away from the susceptible node, then the susceptible node decreases its infected degree by one. Each node from the compartment $S_{k(i+1)}^{U_{na}}$ (resp., $S_{ki}^{U_{na}}$) is shifted to the compartment $S_{ki}^{U_{na}}$ (resp., $S_{k(i-1)}^{U_{na}}$) at rate $(i+1)\omega^U$ (resp., $i\omega^U$). The ninth term depicts the changes of $S_{ki}^{U_{na}}$ owing to their \textit{susceptible degree} (i.e. the number of their susceptible neighbors) increasing by one when the  susceptible nodes are randomly rewired towards by the susceptible endpoint on the $S$-$I$ link after disconnecting from the infected endpoint. Namely, a node from compartment $S_{ki}^{U_{na}}$ jumps to the compartment $S_{(k+1)i}^{U_{na}}$ as it is randomly chosen as the ``new neighbor" in a rewiring event. The average strength of rewiring events is $\bar{\omega}S_I$, where the average rewiring rate $\bar{\omega}$ (see Fig.~\ref{fig2}) is approximated by taking an average over the entire network since we do not know the awareness state of the susceptible endpoint on the $S$-$I$ link that is to be rewired. In the same fashion as in the approximation of $\bar{\beta}$, one has $\bar{\omega}=\omega^A({S^A}/{S})+\omega^U({S^U}/{S})$. Note that a node from the compartment $S_{ki}^{U_{na}}$ (resp., $S_{(k-1)i}^{U_{na}}$) is randomly selected as a new connection target with probability $S_{ki}^{U_{na}}/S$ (resp., $S_{(k-1)i}^{U_{na}}/S$), the event of transition from $S_{ki}^{U_{na}}$ (resp., $S_{(k-1)i}^{U_{na}}$) to $S_{(k+1)i}^{U_{na}}$ (resp., $S_{ki}^{U_{na}}$) occurs at rate $\bar\omega(S_I/S)S_{ki}^{U_{na}}$ (resp., $\bar\omega(S_I/S)S_{(k-1)i}^{U_{na}}$). Explanations for the derivation of Eqs.~(\ref{eq10}) and (\ref{eq11}) are similar except that in the fourth (resp., the second) term on the rhs of Eq.~(\ref{eq10}) [resp., Eq.~(\ref{eq11})], the quantity $H_A=1+U_{AA}/U_A$ (see Fig.~\ref{fig2}) denoting the average number of aware degree of a unaware node reached by following an $A$-$U$ link has been taken into account, and that in the fourth term on the rhs of Eq.~(\ref{eq11}), the variable $G_I=1+S_{II}/S_I$ (see Fig.~\ref{fig2}) measuring the average infected degree of a susceptible reached by following an $I$-$S$ link has been included.

\section{Threshold analysis for the full effective degree model}\label{sec4}

In what follows much attention will be paid on the threshold conditions for disease invasion and awareness information outbreak. In the context of epidemiology, one of the best known quantities that account for the threshold condition for disease invasion is the basic reproduction number $\mathcal{R}_0$, which is defined as the average number of new infections produced by a single infected individual (during its entire infectious period) in a completely susceptible population \cite{AM91,Maia15,LiY18}. Typically, if $\mathcal{R}_0>1$, the disease will break out and finally remains endemic in the population; if $\mathcal{R}_0<1$, the disease will eventually die out.

\begin{figure}
\begin{center}
\includegraphics[width=0.9\columnwidth]{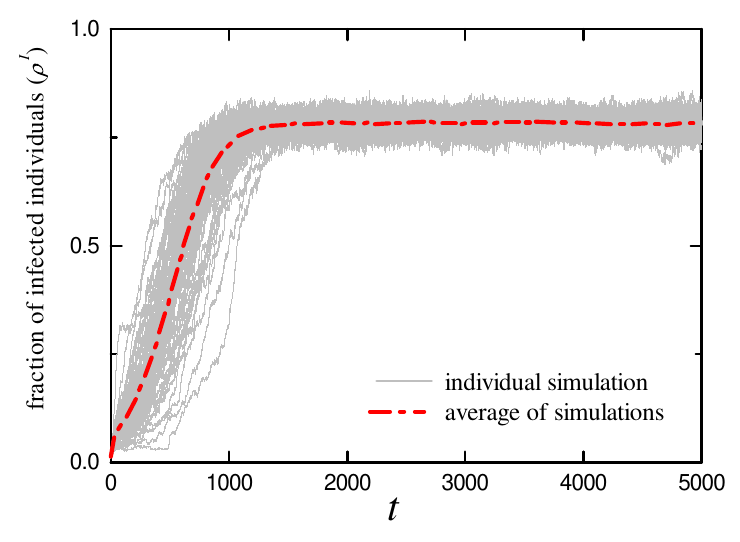}
\caption{(color online) Time series of the density of infected nodes $\rho^{I}$. The simulation results are obtained by running the disease-awareness contagion model with $10\%$ of infected seeds on an initial scale-free multiplex network, in which the disease transmission layer is initially assigned with a power-law degree distribution $P(k)\propto k^{-2.5}$, whilst the awareness information layer is initially assigned a power-law degree distribution $Q(n)\propto n^{-3.5}$, both with degree values ranging from $1$ to $10$. Each grey (thin) curve denotes an individual simulation result and the red (thick) dash-dotted curve marks the average of the $100$ individual simulations. Other parameters are $\alpha=0.04$, $\beta=0.2$, $\mu=0.1$, $\gamma=0.02$, $\omega^A=0.02$, $\omega^U=0.004$.}\label{fig3}
\end{center}
\end{figure}

\textbf{(i) Disease threshold condition}. In the immediate vicinity of the disease-free equilibrium \cite{Maia15,LiY18} for our effective degree model, i.e., as the number of infected individuals is rather small, the possibility for a susceptible individual to have more than one infected neighbors is close to zero. In this case we have $S_{ki}=\sum_{n,a}(S_{ki}^{U_{na}}+S_{ki}^{A_{na}})\approx 0$ for all $i>1$ and
\begin{widetext}
\begin{equation}\label{wsrhofunc}
1+\frac{S_{II}}{S_I}=\frac{\sum_{k=1}^{K}\sum_{i=0}^{k}i^2S_{ki}}{\sum_{k=1}^{K}\sum_{i=0}^{k}iS_{ki}}
=\frac{\sum_{k=1}^{K}S_{k1}}{\sum_{k=1}^{K}S_{k1}}=1.
\end{equation}
\end{widetext}
Therefore, near the disease-free equilibrium the Eq.~(\ref{eq11}) can be written as
\begin{widetext}
\begin{eqnarray}\label{eq13}
\frac{{\rm d} I_{ki}^{A_{na}}}{{\rm d}t}&=&\mu\bigg[(a+1)I_{ki}^{A_{n(a+1)}}-aI_{ki}^{A_{na}}\bigg]+\alpha \bigg(1+\frac{U_{AA}}{U_A}\bigg)\bigg[(n-a+1)I_{ki}^{A_{n(a-1)}}-(n-a)I_{ki}^{A_{na}}\bigg]\nonumber\\
 &&+i\beta\Big(S_{ki}^{U_{na}}+S_{ki}^{A_{na}}\Big)
 +\beta\bigg[(k-i+1)I_{k(i-1)}^{A_{na}}
 -(k-i)I_{ki}^{A_{na}}\bigg]+\gamma\bigg[(i+1)I_{k(i+1)}^{A_{na}}-iI_{ki}^{A_{na}}\bigg]\nonumber\\
 &&-\gamma\*I_{ki}^{A_{na}}+\bigg(\omega^A\frac{S^A}{S}
 +\omega^U\frac{S^U}{S}\bigg)\bigg[(k+1-i)I_{(k+1)i}^{A_{na}}-(k-i)I_{ki}^{A_{na}}\bigg].
\end{eqnarray}
\end{widetext}
It is to be noted that the above approximation is inappropriate when the system is not close to the disease-free equilibrium. In order to obtain the basic reproduction number $\mathcal{R}^{\rm dis}_0$ for disease outbreak from the effective degree model, we collectively linearize the Eqs.~(\ref{eq9}), (\ref{eq10}) and (\ref{eq13}) at the disease-free equilibrium and determine the its stability. In particular, we apply the next generation matrix approach by van den Driessche and Watmough \cite{vW02} to derive the disease basic reproduction number as a spectral radius of matrix ${\bm F}_{d}{\bm V}_{d}^{-1}$, i.e., $\mathcal{R}^{\rm dis}_0=\rho({\bm F}_{d}{\bm V}_{d}^{-1})$, where the matrix ${\bm J}_d={\bm F}_d-{\bm V}_d$ is the Jacobian matrix at the disease-free equilibrium, with the matrix ${\bm F}_d$ accounting for new infections flowing from $S_{k0}^{U_{n0}}$ to $S_{k1}^{U_{n0}}$ and the matrix $-{\bm V}_d$ involving transfers between classes $S_{ki}^{U_{na}}$ for $i\geq 1$, and $S_{ki}^{A_{na}}$, $I_{ki}^{A_{na}}$ for all $k, i, n$ and $a$. Observe from Eq.~(\ref{eq9}) that as new infections individuals in class $S_{k0}^{U_{n0}}$ move into class $S_{k1}^{U_{n0}}$ at a rate
\begin{equation}
\beta\frac{S_{SI}}{S_S}kS_{k0}^{U_{n0}}=\beta\frac{\sum_{k,i}\sum_{n,a}(k-i)i\Big(S_{ki}^{U_{na}}+S_{ki}^{A_{na}}\Big)}{\sum_{k,i}
\sum_{n,a}(k-i)\Big(S_{ki}^{U_{na}}+S_{ki}^{A_{na}}\Big)}kS_{k0}^{U_{n0}},
\end{equation}
and then they can be further transmitted by their infectious neighbors. According to the methodology in \cite{vW02}, the compartments $S_{k0}^{U_{na}}$ and $S_{k0}^{A_{na}}$ do not belong to the disease class, thus they can be disregarded for the threshold and stability calculation. For a fixed $k$ value, in the similar way to \cite{Ma11}, we order the variables with $S_{ki}^{U_{na}}$ equations (in increasing values of $n$) first, and $S_{ki}^{A_{na}}$ equations (in increasing values of $n$) next, then $I_{ki}^{A_{na}}$ equations (in increasing values of $n$) arranged lexicographically, and then in increasing values of $k$. Consequently, the variables are arranged in the order as follows: $S_{11}^{U_{10}}$, $S_{11}^{U_{11}}$, $S_{11}^{U_{20}}$, \dots, $S_{11}^{U_{MM}}$, $S_{11}^{A_{10}}$, \dots, $S_{11}^{A_{MM}}$, $I_{10}^{A_{10}}$, \dots, $I_{10}^{A_{MM}}$, $I_{11}^{A_{10}}$, \dots, $I_{11}^{A_{MM}}$; $S_{21}^{U_{10}}$, \dots, $S_{21}^{U_{MM}}$, $S_{22}^{U_{10}}$, \dots, $S_{22}^{U_{MM}}$, $S_{21}^{A_{10}}$, \dots, $S_{21}^{A_{MM}}$, $S_{22}^{A_{10}}$, \dots, $S_{22}^{A_{MM}}$, $I_{20}^{A_{10}}$, \dots, $I_{20}^{A_{MM}}$, $I_{21}^{A_{10}}$, \dots, $I_{21}^{A_{MM}}$, $S_{22}^{A_{10}}$, \dots, $S_{22}^{A_{MM}}$;$\dots$, $S_{K1}^{U_{10}}$, \dots, $S_{K1}^{U_{MM}}$, \dots, $S_{KK}^{U_{10}}$, \dots, $S_{KK}^{U_{MM}}$, $S_{K1}^{A_{10}}$, \dots, $S_{K1}^{A_{MM}}$, \dots, $S_{KK}^{A_{10}}$, \dots, $S_{KK}^{A_{MM}}$, $I_{K0}^{A_{10}}$, \dots, $I_{K0}^{A_{MM}}$, \dots, $I_{KK}^{A_{10}}$, \dots, $I_{KK}^{A_{MM}}$. With some elementary calculations, ${\bm F}_d$ is shown to be a $\frac{K(3K+5)}{2}\times \frac{K(3K+5)}{2}$ matrix with rank $1$ and can be expressed as
\begin{equation}
{\bm F}_d=\frac{1}{\sum_{k=1}^{K}\sum_{n=1}^{M}kS_{k0}^{U_{n0}}}
\begin{pmatrix}
{\bm X}_1 \\
{\bm X}_2 \\
\vdots\\
{\bm X}_K
\end{pmatrix}
\Big({\bm Y}_1^T,  {\bm Y}_2^T,  \cdots,  {\bm Y}_K^T \Big),
\end{equation}
where ${\bm X}_k$ is a $(3k+1)\times 1$ vector with $kS_{k0}^{U_{n0}}$ in its first entry and zeros elsewhere, and
${\bm Y}_{k}=\begin{pmatrix}
{\bm Y}_k^U\\
~ \\
{\bm Y}_k^A
\end{pmatrix}
$
is a $(3k+1)\times 1$ vector with ${\bm Y}_k^U$ being a $k\times 1$ vector and ${\bm Y}_k^A$ a $(2k+1)\times 1$ vector. For both vectors ${\bm Y}_k^U$ and ${\bm Y}_k^U$, their first $(k-1)$ entries are $(k-1)$, $2(k-2)$, $3(k-3)$, $\dots$, $(k-1)$, while all of their remaining entries are zeros. Moreover, since the degree of each node remains fixed, ${\bm V}_d$ is a block diagonal matrix structured as ${\bm V}_d={\bm V}_{d1}\bigoplus {\bm V}_{d2}\bigoplus\dots\bigoplus {\bm V}_{dK}$, where each block ${\bm V}_{dk}$ ($k=1,2,\cdots,K$) is a $(3k+1)\times(3k+1)$ square matrix depending on $\alpha$, $\mu$, $\beta^U$, $\beta^A$, $\omega^U$, $\omega^A$, $n$ and $k$. Furthermore, it is easy to confirm that for each $k$ all principal minors of ${\bm V}_{dk}$ are positive, implying that the matrix block ${\bm V}_{dk}$ is a nonsingular $M$-matrix \cite{BR97}. Therefore, each inverse matrix ${\bm V}_{dk}^{-1}$ exists and ${\bm V}_{dk}^{-1}\geq 0$, that is, each entry of ${\bm V}_{dk}^{-1}$ is nonnegative \cite{BR97}. As a result, the threshold for disease outbreak is given by
\begin{widetext}
\begin{equation}\label{eq16}
\mathcal{R}^{\rm dis}_{0} = \rho\Big({\bm F}_d {\bm V}_d^{-1}\Big) = \frac{1}{\sum_{k=1}^{K}\sum_{n=1}^{M}kS_{k0}^{U_{n0}}}\sum_{k=1}^{K}\sum_{n=1}^{M}{\bm Y}_k^{T}{\bm V}_{dk}^{-1}{\bm X}_{k}.
\end{equation}
\end{widetext}
If $\mathcal{R}^{\rm dis}_0 < 1$, then the disease-free equilibrium is stable so the disease cannot invade; whereas if $\mathcal{R}^{\rm dis}_0 > 1$, then the disease breaks out and spreads across the population. As shown in Fig.~\ref{fig3}, for a given set of parameters, the average of simulation results shows that the density of infected nodes first increases and finally reaches an steady (endemic) state. It is worth remarking that the threshold transmission rate $\beta_{\rm c}$ for epidemic onset can be derived directly from Eq.~(\ref{eq16}) since $\beta=\beta_{\rm c}$ is equivalent to the critical condition $\mathcal{R}^{\rm dis}_0=1$. In fact, $\beta_{\rm c}=\beta/\mathcal{R}_0^{\rm dis}$. Consequently, $\beta<\beta_{\rm c}$ is equivalent to $\mathcal{R}_0^{\rm dis}<1$, while $\beta>\beta_{\rm c}$ is equivalent to $\mathcal{R}_0^{\rm dis}>1$. Although complicated in form, the formula in Eq.~(\ref{eq16}) will allow for numerical calculations of the transmission threshold for disease onset, which will be demonstrated in Sec.~\ref{sec6}.

\textbf{(ii) Awareness threshold condition}. Similarly to the derivation of the disease threshold condition, in what follows we obtain the awareness basic reproduction number $\mathcal{R}^{\rm awe}_{0}$ as the awareness threshold condition for our effective degree model. Using the similar approximation near the awareness information-free equilibrium (i.e., there is no awareness information at all in the steady state), one has
\begin{widetext}
\begin{eqnarray}\label{eq18}
\frac{{\rm d} S_{ki}^{A_{na}}}{{\rm d}t}&=&-\mu S_{ki}^{A_{na}}+\alpha aS_{ki}^{U_{na}}+\mu\bigg[(a+1)S_{ki}^{A_{n(a+1)}}-aS_{ki}^{A_{na}}\bigg]
 +\alpha\bigg[(n-a+1)S_{ki}^{A_{n(a-1)}}-(n-a)S_{ki}^{A_{na}}\bigg]\nonumber\\
 &&-i\beta S_{ki}^{A_{na}}+\gamma\*I_{ki}^{A_{na}}+\gamma\bigg[(i+1)S_{k(i+1)}^{A_{na}}-iS_{ki}^{A_{na}}\bigg] +\beta\frac{S_{SI}}{S_S}\bigg[(k-i+1)S_{k(i-1)}^{A_{na}}-(k-i)S_{ki}^{A_{na}}\bigg]\nonumber\\
 &&+\omega^{A}\bigg[(i+1)S_{k(i+1)}^{A_{na}}-iS_{ki}^{A_{na}}\bigg]+\bigg(\omega^A \frac{S^A}{S}
 +\omega^U \frac{S^U}{S}\bigg)\frac{S_I}{S}\bigg[S_{(k-1)i}^{A_{na}}-S_{ki}^{A_{na}}\bigg],
\end{eqnarray}
\end{widetext}
\begin{widetext}
\begin{eqnarray}\label{eq19}
\frac{{\rm d} I_{ki}^{A_{na}}}{{\rm d}t}&=&\mu\bigg[(a+1)I_{ki}^{A_{n(a+1)}}-aI_{ki}^{A_{na}}\bigg]
+\alpha\bigg[(n-a+1)I_{ki}^{A_{n(a-1)}}-(n-a)I_{ki}^{A_{na}}\bigg] +i\beta\Big(S_{ki}^{U_{na}}+S_{ki}^{A_{na}}\Big) \nonumber\\
 && +\beta\bigg(1+\frac{S_{II}}{S_I}\bigg)\bigg[(k-i+1)I_{k(i-1)}^{A_{na}}
 -(k-i)I_{ki}^{A_{na}}\bigg] +\gamma\bigg[(i+1)I_{k(i+1)}^{A_{na}}-iI_{ki}^{A_{na}}\bigg] \nonumber\\
 &&-\gamma\*I_{ki}^{A_{na}}+\bigg(\omega^A\frac{S^A}{S}
 +\omega^U\frac{S^U}{S}\bigg)\bigg[(k+1-i)I_{(k+1)i}^{A_{na}}-(k-i)I_{ki}^{A_{na}}\bigg].
\end{eqnarray}
\end{widetext}
Linearizing Eqs.~(\ref{eq9}), (\ref{eq18}) and (\ref{eq19}) at the awareness information-free equilibrium, we proceed as in the derivation of $\mathcal{R}^{\rm dis}_0$ by writing the Jacobian matrix as ${\bm J}_a={\bm F}_a-{\bm V}_a$ and using the same ordering of variables. For our model, matrix ${\bm F}_a$ is shown to be a $\frac{M(3M+7)}{2}\times \frac{M(3M+7)}{2}$ matrix with rank $1$ and can be expressed as
\begin{equation}
{\bm F}_a=\frac{\alpha}{\sum_{n=1}^{M}\sum_{k=1}^{K}nS_{k0}^{U_{n0}}}
\begin{pmatrix}
{\bm W}_1 \\
{\bm W}_2 \\
\vdots\\
{\bm W}_M
\end{pmatrix}
\Big({\bm Z}_1^T,  {\bm Z}_2^T,  \cdots,  {\bm Z}_M^T \Big),
\end{equation}
in which ${\bm W}_n$ is a $(3n+2)\times 1$ vector with $nS_{k0}^{U_{n0}}$ in its first entry and zeros elsewhere, and
${\bm Z}_{k}$ is also a $(3n+2)\times 1$ vector with its first $(n-1)$ entries equal to $(n-1)$, $2(n-2)$, $3(n-3)$, $\dots$, $(n-1)$, and all its remaining entries are zeros. Moreover, matrix ${\bm V}_a$ is a block diagonal matrix structured as ${\bm V}_a={\bm V}_{a1}\bigoplus {\bm V}_{a2}\bigoplus\dots\bigoplus {\bm V}_{aM}$, of which each block ${\bm V}_{an}$ ($n=1,2,\cdots,M$) is a $(3n+2)\times(3n+2)$ nonsingular $M$-matrix \cite{BR97}. Therefore, the awareness threshold condition is given by
\begin{widetext}
\begin{equation}\label{eq22}
\mathcal{R}^{\rm awe}_{0} = \rho\Big({\bm F}_a {\bm V}_a^{-1}\Big) = \frac{\alpha}{\sum_{n=1}^{M}\sum_{k=1}^{K}nS_{k0}^{U_{n0}}}\sum_{n=1}^{M}\sum_{k=1}^{K}{\bm Z}_n^{T}{\bm V}_{an}^{-1}{\bm W}_{n}.
\end{equation}
\end{widetext}
If $\mathcal{R}^{\rm awe}_0 < 1$, then the awareness information-free equilibrium is stable so the awareness information cannot invade and in the end all individuals are unaware of the disease information; whereas if $\mathcal{R}^{\rm awe}_0 > 1$, then the awareness information breaks out and eventually saturates in the population. In an analogous manner to the deduction of $\beta_{\rm c}$, the critical diffusion rate $\alpha_{\rm c}$ for awareness information outbreak can be derived as $\alpha_{\rm c}=\alpha/\mathcal{R}_0^{\rm awe}$. Therefore, $\alpha>\alpha_{\rm c}$ is equivalent to $\mathcal{R}_0^{\rm awe}>1$, whereas $\alpha<\alpha_{\rm c}$ is equivalent to $\mathcal{R}_0^{\rm awe}<1$.

\section{Disease reproduction number in a reduced effective degree model}\label{sec5}
Similar to Ref.~\cite{Ma11}, in this section we consider a reduced SIS-UAU effective degree model in which the term $\beta(k-i+1)I_{k(i-1)}^{A_{na}}$ in Eq.~(\ref{eq13}) is set to zero. In this case the differential equations for $I_{k0}^{A_{na}}$ are decoupled from the other equations and hence they can be ignored from the dynamical system for threshold and stability analysis. In this regard, we only need to consider a reduced system of order $\Big(\sum_{k=1}^{K}3k\Big)=3K(K+1)/2$. Rearrange the order of variables as follows: $S_{11}^{U_{na}}$, $S_{11}^{A_{na}}$, $I_{11}^{A_{na}}$, $S_{21}^{U_{na}}$, $S_{21}^{A_{na}}$, $I_{21}^{A_{na}}$, $S_{22}^{U_{na}}$, $S_{22}^{A_{na}}$, $I_{22}^{A_{na}}$, $\dots$, $S_{KK}^{U_{na}}$, $S_{KK}^{A_{na}}$, $I_{KK}^{A_{na}}$. It is to be noted that each of the above variables is itself ordered in increasing value of $n$; for example, $S_{11}^{U_{na}}$ is referred to as $\big(S_{11}^{U_{10}}, S_{11}^{U_{11}}, S_{11}^{U_{20}}, \dots, S_{11}^{U_{MM}}\big)$. Note that ${\bm V}_d={\bm V}_{d1}\bigoplus {\bm V}_{d2}\bigoplus\dots\bigoplus {\bm V}_{dK}$, where ${\bm V}_{dk}$ $(k=1,2,\cdots,K)$ are block matrices of ${\bm V}_d$. In this reduced effective degree model, the blocks ${\bm V}_{dk}$ now turn to be upper triangular with diagonal blocks of order $3$. Due to the sparsity of matrix ${\bm F}_{d}$, only the blocks on variables $S_{k1}^{U_{na}}$, $S_{k1}^{A_{na}}$, $I_{k1}^{A_{na}}$ contribute to matrix ${\bm F}_{d}{\bm V}_{d}^{-1}$. Consequently, the block matrices of ${\bm V}_{d}$ are
\begin{equation}\label{eq23}
{\bm V}_{dk} =
\begin{pmatrix}
f_1 & -\mu & 0\\
0 & f_2 & -\gamma\\
-\beta & -\beta & f_3
\end{pmatrix},\qquad k=1,2,\cdots,K
\end{equation}
with inverse
\begin{equation}\label{eq24}
{\bm V}_{dk}^{-1} =
\frac{1}{\gamma\beta(\mu+f_1)-f_1f_2f_3}\begin{pmatrix}
\gamma\beta-f_{2}f_{3} & -\mu f_3 & -\mu\gamma\\
-\gamma\beta & -f_{1}f_{3} & -\gamma f_1\\
-\beta f_2 & -\beta(\mu+f_1) & -f_{1}f_{2}
\end{pmatrix},
\end{equation}
where $f_1=\beta+\gamma+\omega^{U}$, $f_2=n\alpha+\mu+\beta+\gamma+\omega^{A}$, and $f_3=n\alpha+2\gamma+(k-1)(\beta+\omega^{U})$.
In addition, the corresponding blocks of ${\bm F}_{d}$ are
\begin{equation}
{\bm F}_{dk}=\frac{1}{\sum_{k=1}^{K}\sum_{n=1}^{M}kS_{k0}^{U_{n0}}}
\begin{pmatrix}
k(k-1)\beta S_{k0}^{U_{n0}} & k(k-1)\beta S_{k0}^{U_{n0}} & 0\\
0 & 0 & 0 \\
0 & 0 & 0
\end{pmatrix}.
\end{equation}
As a result, in the reduced effective degree model the basic reproduction number $\mathcal{R}^{\rm dis}_0=\rho({\bm F}_{d}{\bm V}_{d}^{-1})$ reads
\begin{equation}\label{eq26}
\mathcal{R}^{\rm dis}_0=\sum_{k=1}^{K}\sum_{n=1}^{M}\frac{\beta f_2f_3 S_{k0}^{U_{n0}}}{f_1f_2f_3-\gamma\beta(\mu+f_1)}\frac{k(k-1)}{\sum_{k=1}^{K}\sum_{n=1}^{M}kS_{k0}^{U_{n0}}}
\end{equation}
It is easy to check that in the denominator $f_1f_2f_3-\gamma\beta (\mu+f_1)>0$, implying $f_1f_2f_3>f_1f_2f_3-\gamma\beta (\mu+f_1)>0$. So, we have a lower bound for the disease reproduction number, that is
\begin{equation}\label{eq27}
\mathcal{R}^{\rm dis}_0>\sum_{k=1}^{K}\sum_{n=1}^{M}\frac{\beta S_{k0}^{U_{n0}}}{f_1}\frac{k(k-1)}{\sum_{k=1}^{K}\sum_{n=1}^{M}kS_{k0}^{U_{n0}}}=\frac{\beta}{\beta+\gamma+\omega^{U}}\frac{\langle k(k-1)\rangle}{\langle k\rangle} \triangleq {\mathcal{R}}_0^{\rm LB},
\end{equation}
where $\langle \cdot \rangle$ denotes the expectation calculated from the given degree distribution $P(k)=S_{k0}=\sum_{n=1}^{M}S_{k0}^{U_{n0}}$.

To compare our reduced model with the original SIS-UAU effective degree model given by Eqs.~(\ref{eq9})-(\ref{eq11}), observe that every entry of ${\bm V}_{dk}$ in the original effective degree model is smaller than or equal to the corresponding entry of ${\bm V}_{dk}$ in the reduced model. Since ${\bm V}_{dk}$ is a nonsingular $M$-matrix, by the theory of $M$-matrix \cite{BR97}, we can argue that each entry of ${\bm V}_{dk}^{-1}$ in the original effective degree model is greater than or equal to the corresponding ${\bm V}_{dk}^{-1}$ entry in the reduced model. Note that matrix ${\bm F}_d {\bm V}_d^{-1}$ is nonnegative, so the basic reproduction number $\mathcal{R}_0^{\rm dis}$ for the original effective degree model is larger than or equal to $\mathcal{R}_0^{\rm dis}$ for the reduced model. This implies that for the same parameters $\beta$, $\gamma$ and $\omega^{U}$ and the degree distribution $P(k)$, the disease basic reproduction number $\mathcal{R}_0^{\rm dis}$ of the effective degree model always has a lower bound, i.e., $\mathcal{R}_0^{\rm dis}>{\mathcal{R}}_0^{\rm LB}=\frac{\beta}{\beta+\gamma+\omega^{U}}\frac{\langle k(k-1)\rangle}{\langle k\rangle}$.

\textbf{(i) Effects of link rewiring on disease outbreak.} Consider the case of no rewiring, that is, when rewiring parameters $\omega^{U}$ and $\omega^{A}$ are set to zero, the basic reproduction number for disease outbreak turns to
\begin{eqnarray}
\widetilde{\mathcal{R}}_0^{\rm dis} = \rho(\widetilde{{\bm F}}_{d}\widetilde{{\bm V}}_{d}^{-1})
&=& \sum_{k=1}^{K}\sum_{n=1}^{M}\frac{\beta\widetilde{f}_2\widetilde{f}_3 S_{k0}^{U_{n0}}}{\widetilde{f}_1\widetilde{f}_2\widetilde{f}_3-\gamma\beta(\mu+\widetilde{f}_1)}
\frac{k(k-1)}{\sum_{k=1}^{K}\sum_{n=1}^{M}kS_{k0}^{U_{n0}}}\nonumber\\
&=& \sum_{k=1}^{K}\sum_{n=1}^{M}\frac{\beta\widetilde{f}_2\widetilde{f}_3 S_{k0}^{U_{n0}}}{\widetilde{f}_1\widetilde{f}_2\widetilde{f}_3-\gamma\beta(\mu+\widetilde{f}_1)}
\frac{k(k-1)}{\langle k\rangle},
\end{eqnarray}
where $\widetilde{f}_1=\beta+\gamma$, $\widetilde{f}_2=n\alpha+\mu+\beta+\gamma$, and $\widetilde{f}_3=n\alpha+2\gamma+(k-1)\beta$.

As presented in \textbf{Appendix A}, the disease basic reproduction number $\mathcal{R}_0^{\rm dis}$ in our effective degree model with rewiring is proven to be smaller as compared with the model without rewiring, i.e., $\mathcal{R}_0^{\rm dis} < \widetilde{\mathcal{R}}_0^{\rm dis}$. This result implies that the adaptive link rewiring scheme plays an important role in preventing the disease outbreak by decreasing the basic reproduction number $\mathcal{R}_0^{\rm dis}$.

\textbf{(ii) Special case---the model without awareness contagion.} Consider the special case when the awareness layer is absent, namely, there is no awareness information propagating among the population, then the original two-layer network effective degree model is degenerated into a single-layer network effective degree model (which was studied numerically by Marceau \textit{et al.} \cite{Marceau10}, where the theoretical disease outbreak threshold was not provided). In addition, each node only has one of the two states: susceptible or infected, and does not possess the awareness information states. Moreover, there is only one parameter $\omega$ for link rewiring for all susceptible nodes (i.e., $\omega^U=\omega^A=\omega$) and there is no parameters for information contagion (i.e., $\alpha=\mu=0$). Thus, in this case, the basic reproduction number for disease outbreak becomes
\begin{widetext}
\begin{equation}\label{eq32}
\widehat{\mathcal{R}}_0^{\rm dis}(\omega)=\sum_{k=1}^{K}\sum_{n=1}^{M}\frac{\beta\widehat{f}_2\widehat{f}_3S_{k0}^{U_{n0}}}{\widehat{f}_1
\widehat{f}_2\widehat{f}_3-\gamma\beta\widehat{f}_1} \frac{k(k-1)}{\langle k\rangle},
\end{equation}
\end{widetext}
where $\widehat{f}_1=\widehat{f}_2=\beta+\gamma+\omega$, and $\widehat{f}_3=2\gamma+(k-1)(\beta+\omega)$. By substituting $\widehat{f}_1$, $\widehat{f}_2$ and $\widehat{f}_3$ into Eq.~(\ref{eq32}) it is easy to obtain
\begin{widetext}
\begin{eqnarray}
\widehat{\mathcal{R}}_0^{\rm dis}(\omega) &=& \sum_{k=1}^{K}\sum_{n=1}^{M}\frac{\beta\widehat{f}_3S_{k0}^{U_{n0}}}{
\widehat{f}_2\widehat{f}_3-\gamma\beta} \frac{k(k-1)}{\langle k\rangle} = \sum_{k=1}^{K}\frac{\beta\widehat{f}_3P(k)}{
\widehat{f}_2\widehat{f}_3-\gamma\beta} \frac{k(k-1)}{\langle k\rangle}, \nonumber\\
&=& \sum_{k=1}^{K}\frac{\beta\Big[2\gamma+(k-1)(\beta+\omega)\Big]}{
(\beta+\gamma+\omega)\Big[2\gamma+(k-1)(\beta+\omega)\Big]-\gamma\beta} \frac{k(k-1)P(k)}{\langle k\rangle}.
\end{eqnarray}
\end{widetext}

As the rewiring process is removed from the model, i.e., if $\omega=0$, then the basic reproduction number turns into
\begin{equation}
\widehat{\mathcal{R}}_0^{\rm dis}(0) = \sum_{k=1}^{K}\frac{2\beta\gamma+(k-1)\beta^2}{2\gamma^2+(k-1)\beta^2+k\beta\gamma}\frac{k(k-1)P(k)}{\langle k\rangle},
\end{equation}
which reproduces the result obtained by \textit{Lindquist et al.} (see Appendix C in Ref. \cite{Ma11}).
Moreover, since $k\geq 1$, it can be easily verified that for any positive rewiring rate $\omega$, the following inequality holds:
\begin{equation}
\frac{2\beta\gamma+(k-1)\beta^2}{2\gamma^2+(k-1)\beta^2+k\beta\gamma} >
\frac{\beta\Big[2\gamma+(k-1)(\beta+\omega)\Big]}{
(\beta+\gamma+\omega)\Big[2\gamma+(k-1)(\beta+\omega)\Big]-\gamma\beta}.
\end{equation}
Therefore, one has for $\omega>0$
\begin{equation}
\widehat{\mathcal{R}}_0^{\rm dis}(0)>\widehat{\mathcal{R}}_0^{\rm dis}(w),
\end{equation}
which highlights the important role of the adaptive rewiring in decreasing the disease reproduction number and hence hampering the disease outbreak.

\begin{figure}[ht]
\begin{tabular}{cc}
  \mbox{\includegraphics*[width=0.45\columnwidth,clip=0]{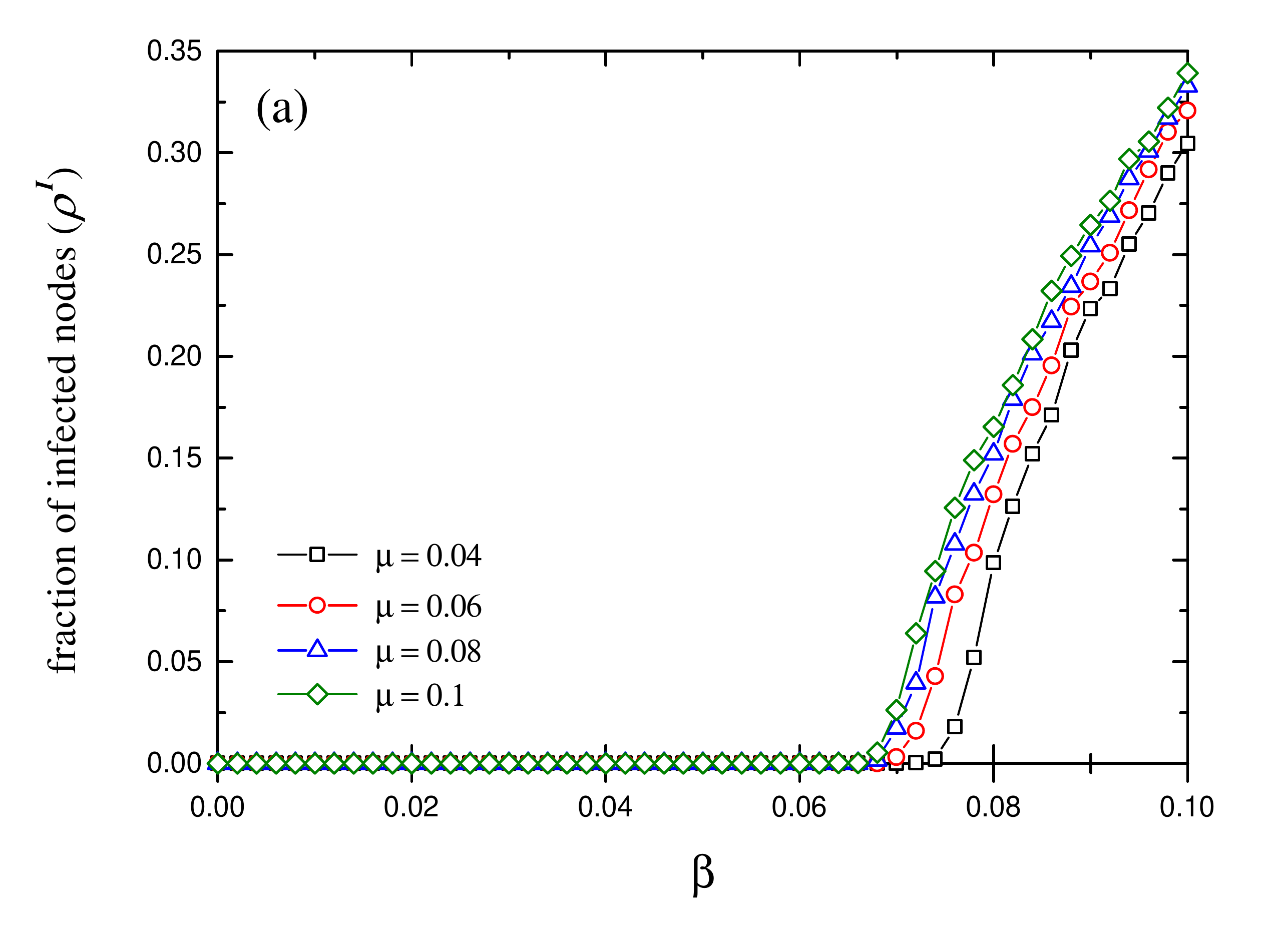}} &
  \mbox{\includegraphics*[width=0.45\columnwidth,clip=0]{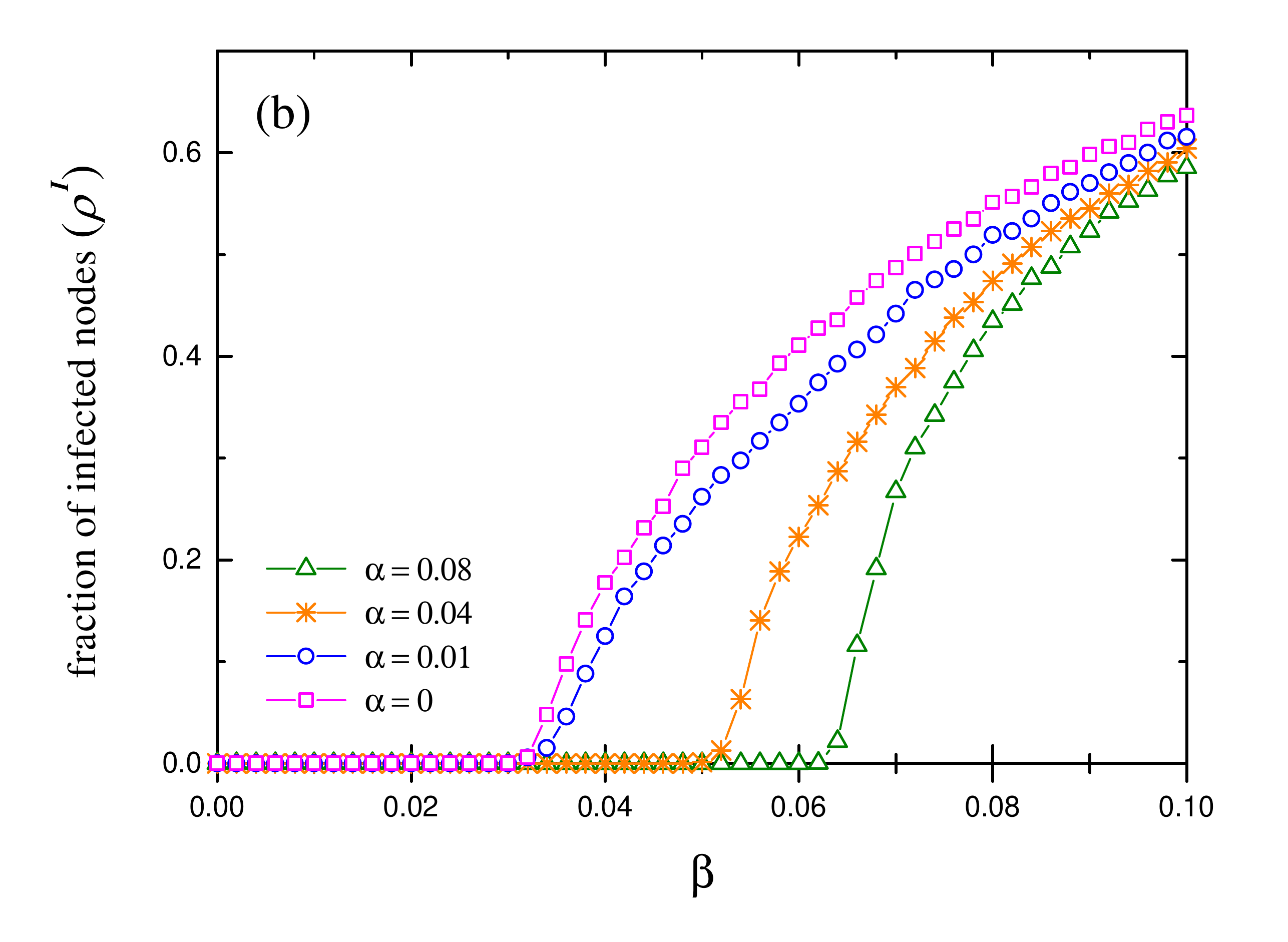}}
  \\
  \mbox{\includegraphics*[width=0.45\columnwidth,clip=0]{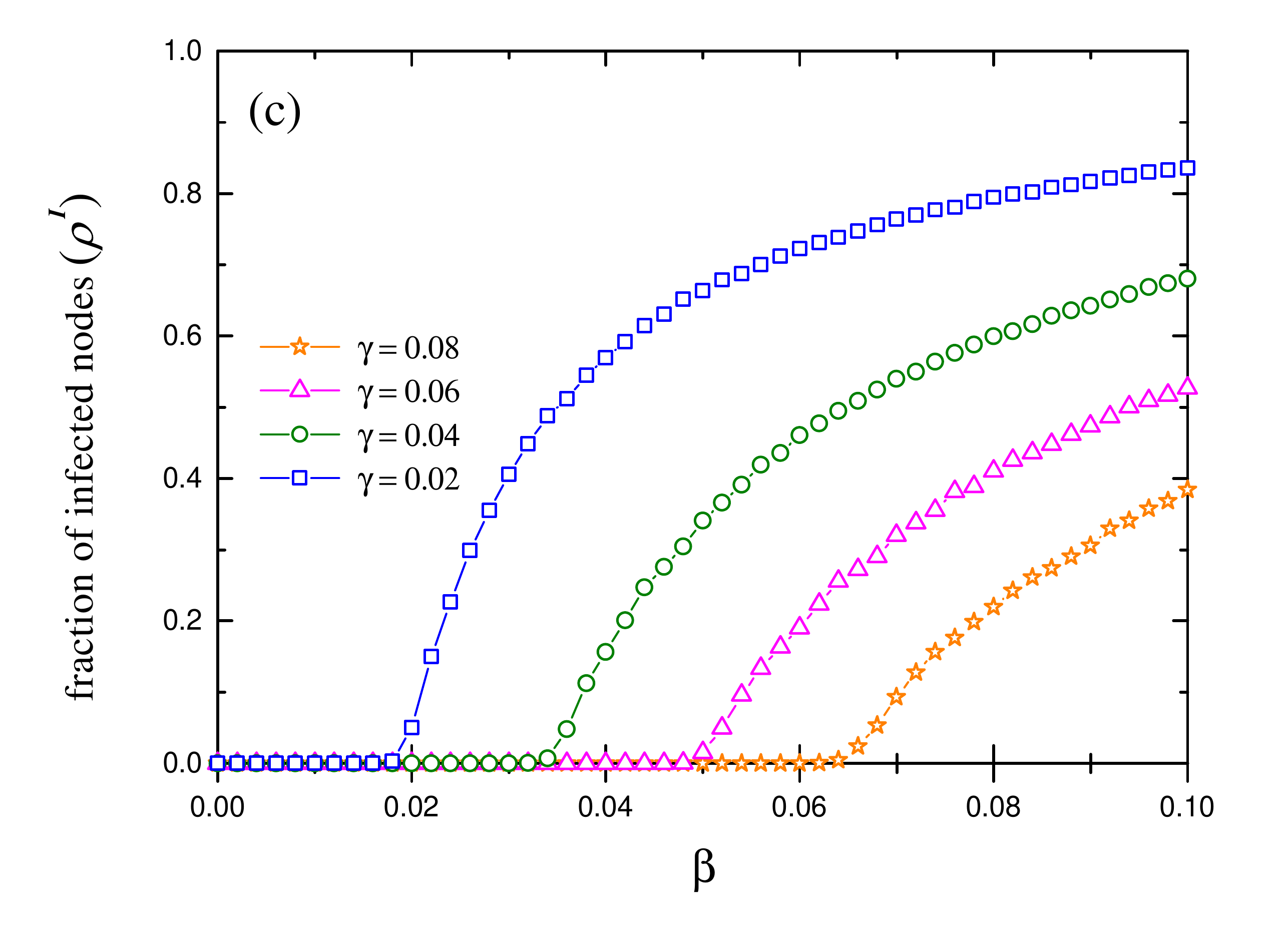}} &
  \mbox{\includegraphics*[width=0.45\columnwidth,clip=0]{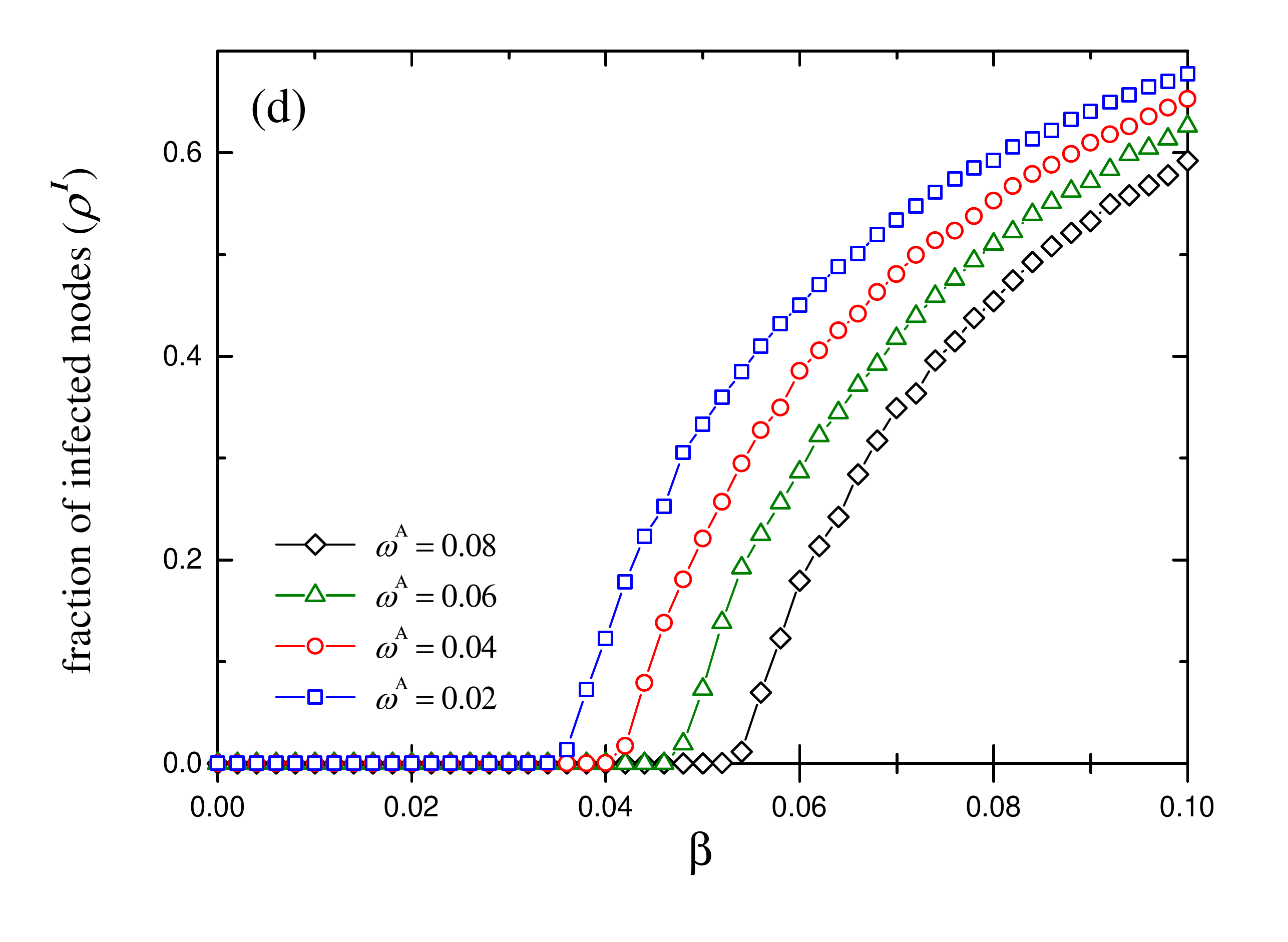}}
\end{tabular}
\caption{(color online) The fraction $\rho^I$ of infected nodes in the stationary state as a function of the disease transmission rate $\beta$ for different values of (a) the awareness information forgetting rate $\mu$, (b) the awareness information spreading rate $\alpha$, (c) the recovery rate $\gamma$ from disease, and (d) the rewiring rate $\omega^A$ for aware susceptible individuals. In this plot, the results are obtained on an initial Poisson multiplex network, in which the disease transmission layer is initially assigned with a Poisson degree distribution $P(k)\propto 2^{k}e^{-2}/k!$, whilst the awareness information layer is initially assigned a power-law degree distribution $Q(n)\propto 3^{n}e^{-3}/n!$, both with degree values ranging from $1$ to $10$. Each point is obtained by averaging over $200$ independent realizations. Other parameters are set in panels (a) $\alpha=0.02$, $\gamma=0.08$, $\omega^A=0.05$, $\omega^U=0.01$; (b) $\gamma=0.04$, $\mu=0.06$, $\omega^A=0.08$, $\omega^U=0.0016$; (c) $\alpha=0.02$, $\mu=0.06$, $\omega^A=0.02$, $\omega^U=0.004$; (d) $\alpha=0.02$, $\gamma=0.04$, $\mu=0.04$, $y=\omega^U/\omega^A=0.2$.}\label{fig4}
\end{figure}

\begin{figure}[ht]
\begin{tabular}{cc}
  \mbox{\includegraphics*[width=0.45\columnwidth,clip=0]{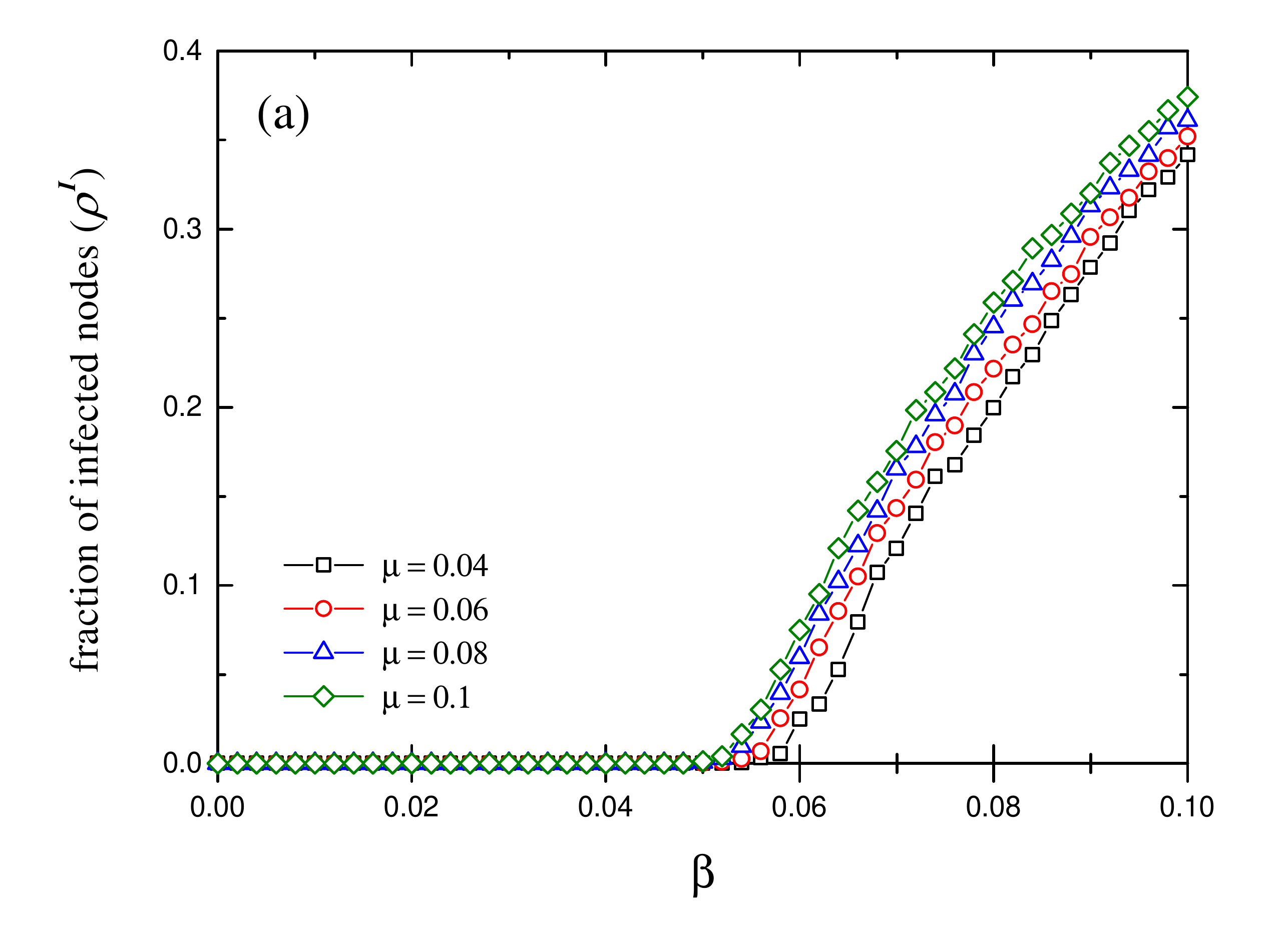}} &
  \mbox{\includegraphics*[width=0.45\columnwidth,clip=0]{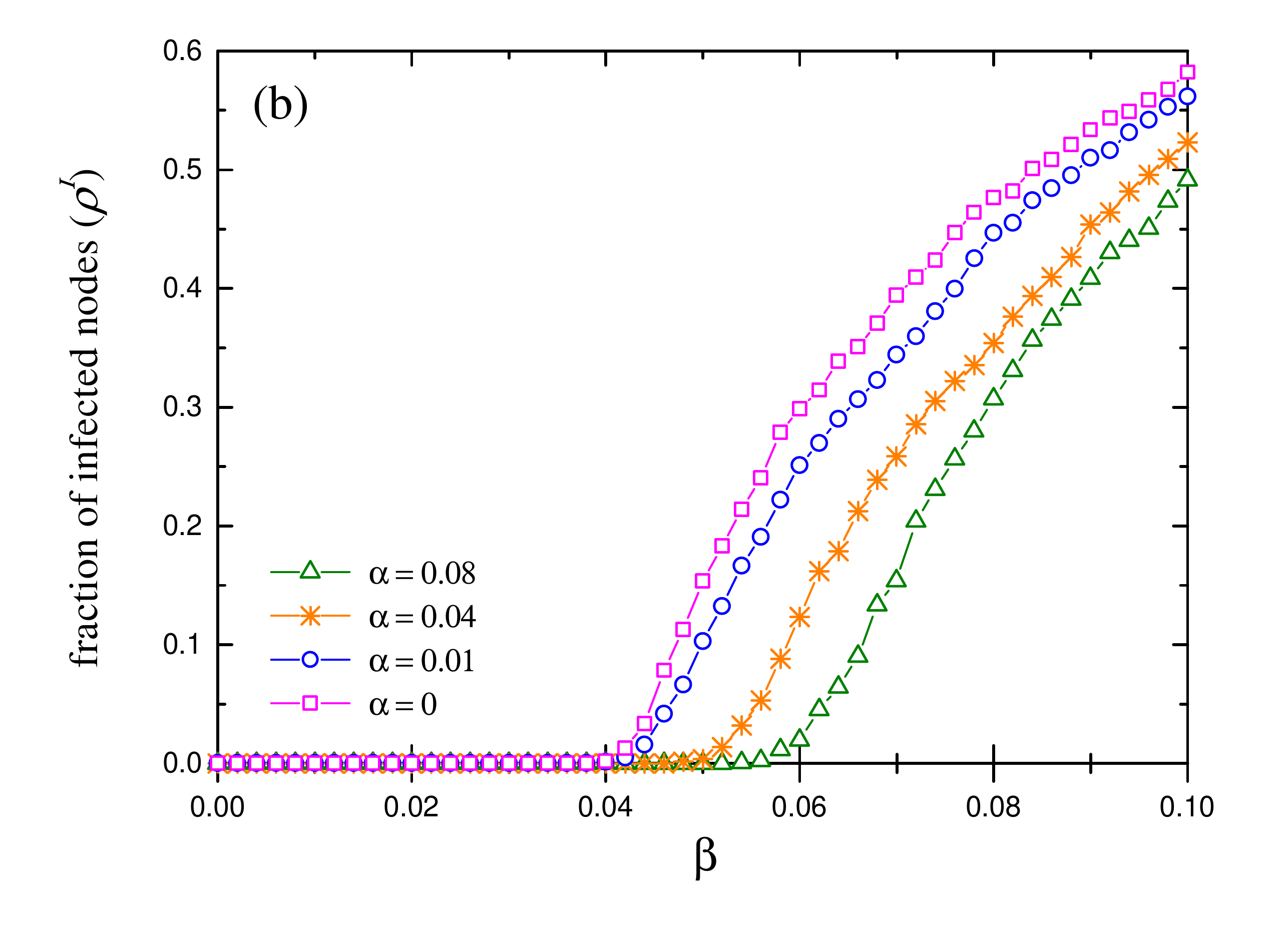}}
  \\
  \mbox{\includegraphics*[width=0.45\columnwidth,clip=0]{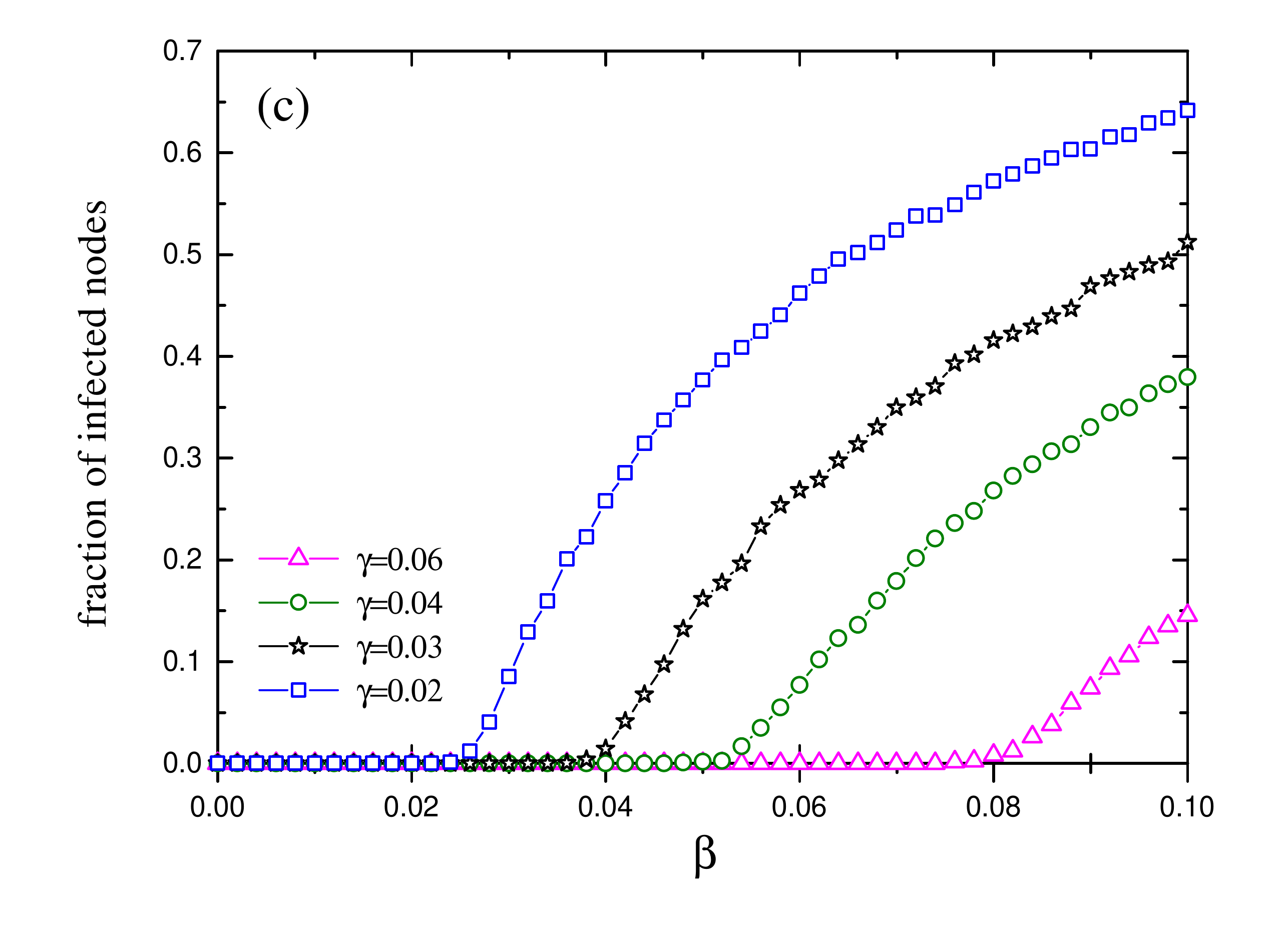}} &
  \mbox{\includegraphics*[width=0.45\columnwidth,clip=0]{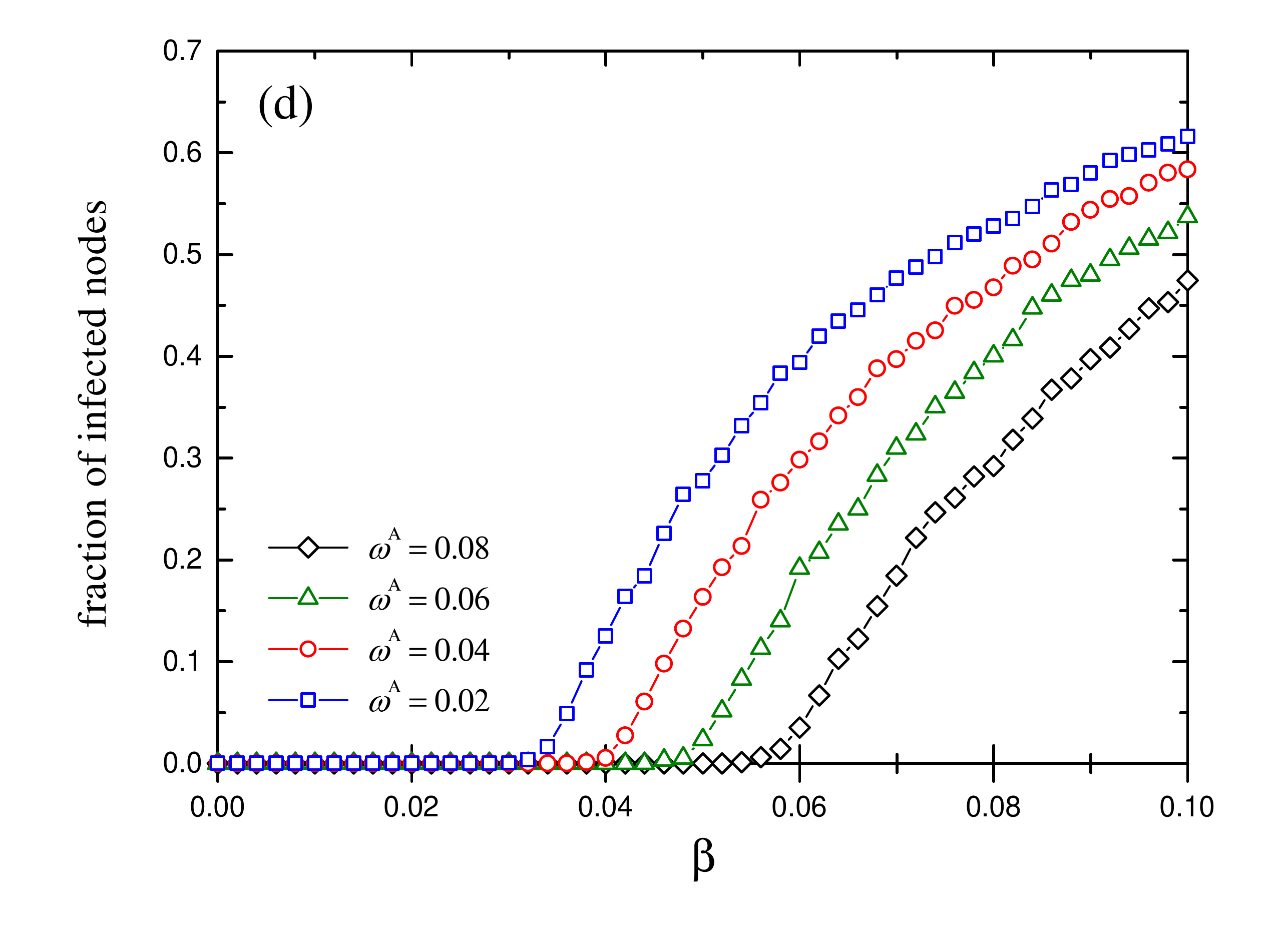}}
\end{tabular}
\caption{(color online) The fraction $\rho^I$ of infected nodes in the stationary state as a function of the disease transmission rate $\beta$ for different values of (a) the awareness information forgetting rate $\mu$, (b) the awareness information spreading rate $\alpha$, (c) the recovery rate $\gamma$ from disease, and (d) the rewiring rate $\omega^A$ for aware susceptible individuals. In this plot, the results are obtained on an initial scale-free multiplex network, in which the disease transmission layer is initially assigned with a power-law degree distribution $P(k)\propto k^{-2.5}$, whilst the awareness information layer is initially assigned a power-law degree distribution $Q(n)\propto n^{-3.5}$, both with degree values ranging from $1$ to $10$. Each point is obtained by averaging over $200$ independent realizations. Other parameters are set in panels (a) $\alpha=0.08$, $\gamma=0.04$, $\omega^A=0.02$, $\omega^U=0.0004$; (b) $\gamma=0.02$, $\mu=0.02$, $\omega^A=0.05$, $\omega^U=0.01$; (c) $\alpha=0.02$, $\mu=0.06$, $\omega^A=0.02$, $\omega^U=0.0004$; (d) $\alpha=0.04$, $\gamma=0.02$, $\mu=0.04$, $y=\omega^U/\omega^A=0.2$.}\label{fig5}
\end{figure}

\section{Simulation results} \label{sec6}
In this section, numerical computation and stochastic simulation have been carried out to verify the theoretical analysis for our effective degree model. In the simulation, the network size (i.e., the number of nodes) is set to $N=10^4$. Initially, all nodes are set to be susceptible and unaware of disease information, and then $10\%$ of nodes are randomly chosen as the infected seeds that will trigger the contagion process across the network in both layers.

\begin{figure}[ht]
\begin{tabular}{cc}
  \mbox{\includegraphics*[width=0.45\columnwidth,clip=0]{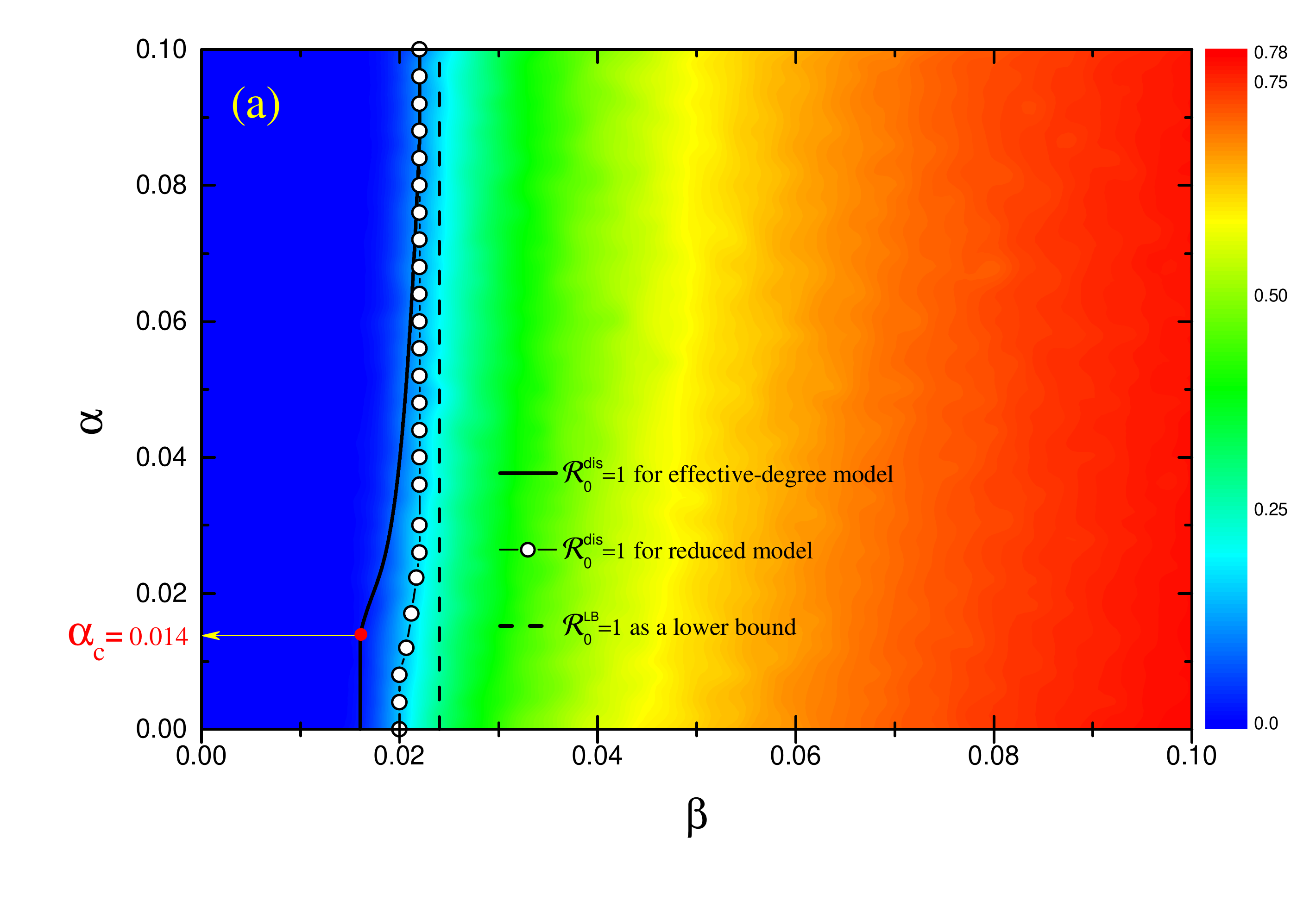}} &
  \mbox{\includegraphics*[width=0.45\columnwidth,clip=0]{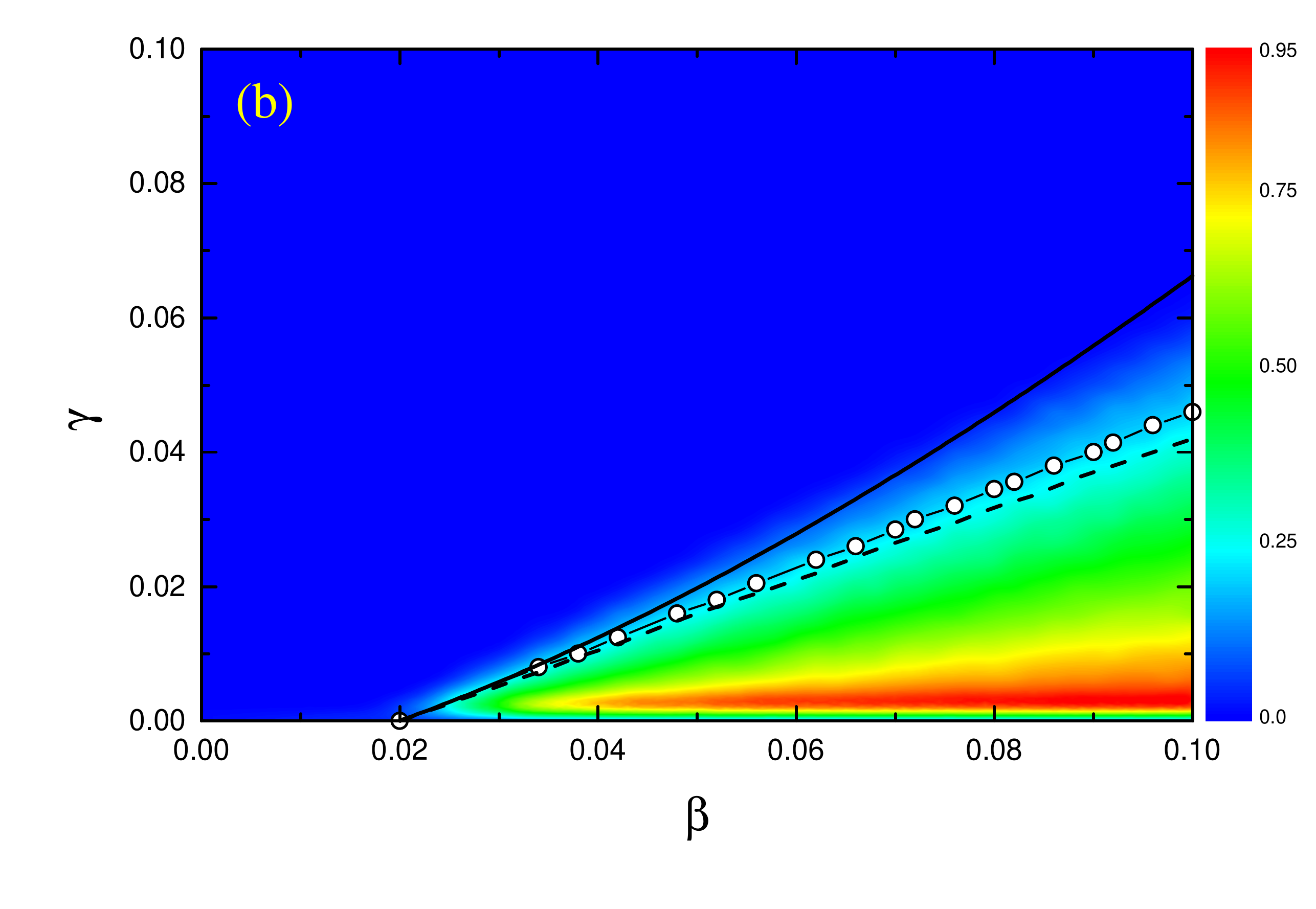}}
  \\
  \mbox{\includegraphics*[width=0.45\columnwidth,clip=0]{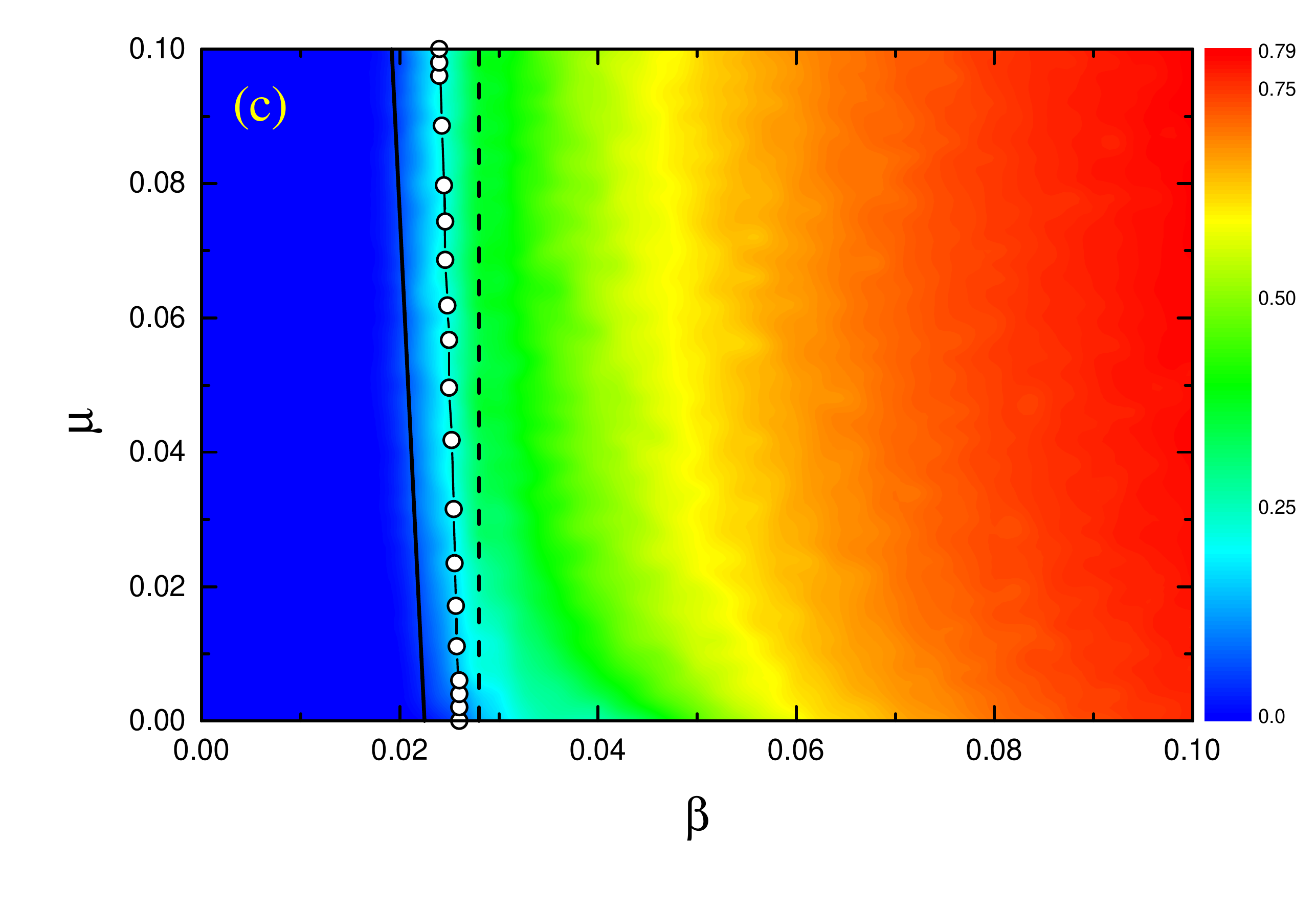}} &
  \mbox{\includegraphics*[width=0.45\columnwidth,clip=0]{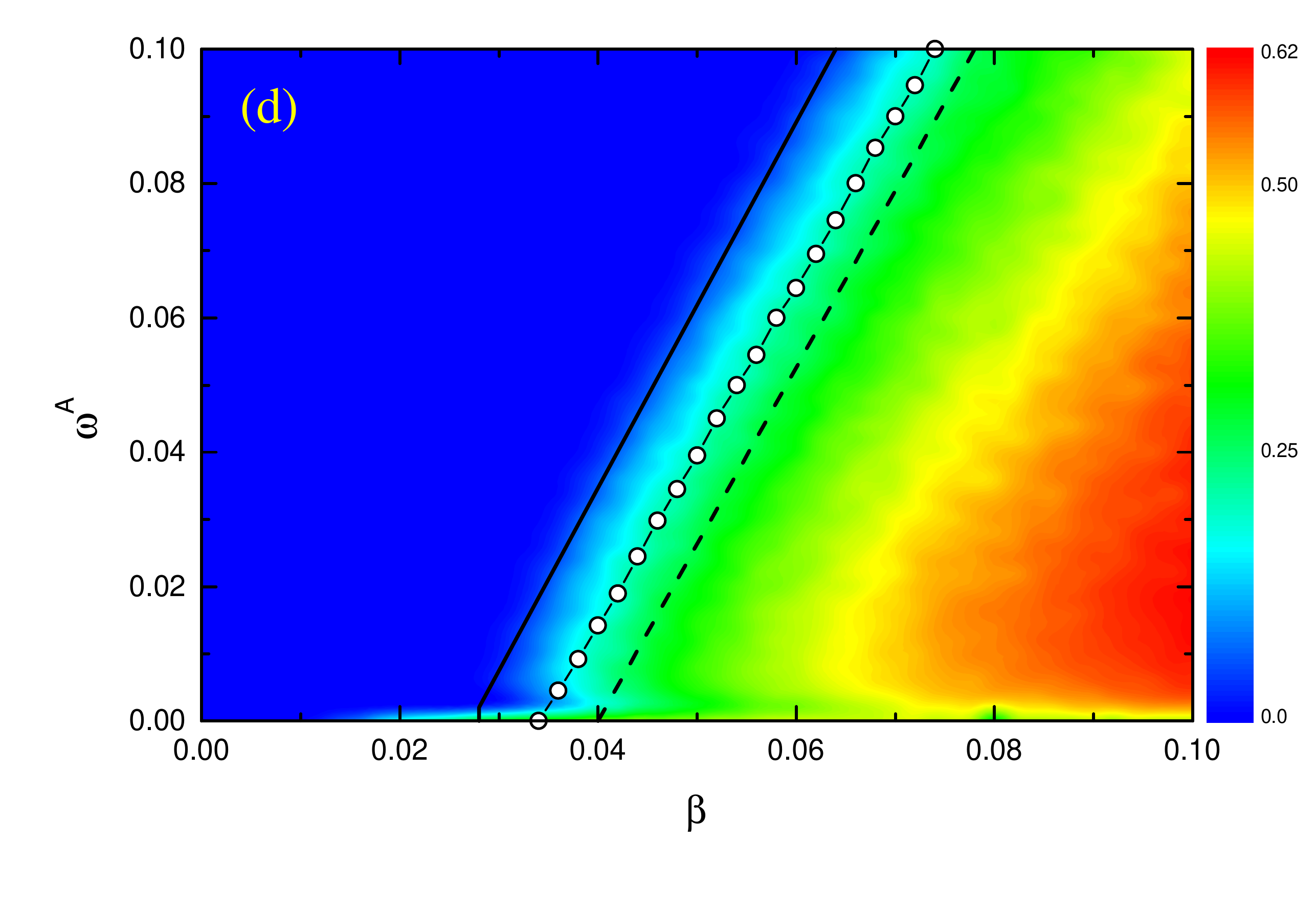}}
\end{tabular}
\caption{(color online) The fraction $\rho^I$ (of which the simulation results are represented by the color bar in the heat map) of infected individuals in the stationary state as a function of $\beta$ and (a) $\alpha$, (b) $\gamma$, (c) $\mu$, and (d) $\omega^A$. All full phase diagram are obtained for the same initial multiplex network structure as in Fig.~\ref{fig5}. For comparison, the theoretical threshold conditions for disease outbreak obtained from our effective degree model [Eq.~(\ref{eq16})], the reduced model [Eq.~(\ref{eq26})], and the lower bound [Eq.~(\ref{eq27})] are indicated by solid curves, empty circles, and dashed curves, respectively. The red filled circle in panel (a) marks the metacritical point $(\alpha_{\rm c}, \beta_{\rm c})$. Note that the threshold transmission rate for disease outbreak is $\beta_{\rm c}=\beta/\mathcal{R}_0^{\rm dis}$, above which the disease breaks out. Analogously, the threshold diffusion rate for awareness information outbreak is $\alpha_{\rm c}=\alpha/\mathcal{R}_0^{\rm awe}$, above which the awareness information prevails. The simulation results are obtained by averaging over $100$ independent realizations. Other parameters are used for panels: (a) $\mu=0.02$, $\gamma=0.01$, $\omega^A=0.01$, $\omega^U=0.002$; (b) $\alpha=0.08$, $\mu=0.09$, $\omega^A=0.05$, $\omega^U=0.01$; (c) $\alpha=0.02$, $\gamma=0.01$, $\omega^A=0.02$, $\omega^U=0.004$; (d) $\alpha=0.04$, $\mu=0.06$, $\gamma=0.02$, $y=\omega^U/\omega^A=0.2$.}\label{fig6}
\end{figure}

In Fig.~\ref{fig3}, we present the results of $100$ stochastic simulations for the temporal evolution of the fraction of infected nodes on an artificial multiplex networks with both layers initially set as scale-free. Each grey solid curve represents an independent simulation trajectory, while the red dash-dotted curve is the average simulation result. It is illustrated that although there are evident fluctuations in the time required to reach the expansion phase and the final size (which is largely due to the impact of random events in the early stage) , the proportion of infected nodes eventually reaches the stationary state, meaning the disease becomes endemic.

To further investigate the effects of parameters on disease transmission, we report the density $\rho^I$ of infected nodes [given by Eq.~(\ref{eq5})] as a function of the disease transmission rate $\beta$ with different values of other parameters in an initial multiplex network with Poisson degree distributions and power-law degree distributions in Fig.~\ref{fig4} and Fig.~\ref{fig5}, respectively. Direct comparison between the results in Fig.~\ref{fig4} and Fig.~\ref{fig5} reveals that the initial network structures have little impact on the qualitative behavior of the $\rho^I$ versus $\beta$ bifurcation dynamics. It is worth emphasizing that the effect of initial network setting can be ignored since the process of link rewiring causes the network to converge to an randomly connected network with a Poisson degree distribution \cite{Gross06,Peng16}. As shown in Fig.~\ref{fig9} in Appendix B, in the case of adaptive rewiring, the resulting network will converge to a narrowly distributed degree distribution even if the initial network setting is a perfect power law. Therefore, from now on, in our simulations we will simply apply an initial scale-free network setting just as done in Refs.~\cite{Gran13,Gran14,Zhang17}. As shown in Fig.~\ref{fig4}(a) and Fig.~\ref{fig5}(a), with the increase of the information forgetting rate $\mu$ from $0.04$ to $0.1$, the disease outbreak threshold decreases slightly and the value of $\rho^I$ rises correspondingly when the disease breaks out. This suggests that the disease outbreak threshold is not much sensitive (i.e., it is relatively robust) to the information forgetting rate. As illustrated in Fig.~\ref{fig4}(b) and Fig.~\ref{fig5}(b), with the increase of the information propagating rate $\alpha$, the disease prevalence $\rho^I$ increases accordingly when the disease breaks out. However, the disease outbreak threshold almost remains unchanged for the small values of $\alpha=0$ and $\alpha=0.01$; however, as $\alpha$ becomes large enough (in this case $\alpha=0.04$ and $\alpha=0.08$), the disease threshold grows evidently with $\alpha$. This implies a possible critical phenomenon in the dependence of the disease threshold $\beta_{\rm c}$ on the information spreading rate $\alpha$. That is, there exists a critical value $\alpha_{\rm c}$ such that only when $\alpha<\alpha_{\rm c}$ will $\beta_{\rm c}$ grow with $\alpha$, otherwise $\beta_{\rm c}$ keeps unchanged. This meta-critical behavior is further verified in Fig.\ref{fig6}(a). The enhancement of disease information propagation would help reduce the infectious disease prevalence to some extent; nevertheless, from the perspective of disease prevention, only a large enough information diffusion rate will help enlarge the disease threshold to curb the onset of disease. In Fig.~\ref{fig4}(c) and Fig.~\ref{fig5}(c), it is shown that, as expected, increasing the recovery rate $\gamma$ from disease infection will help impede the disease transmission by enlarging the disease threshold as well as reducing the epidemic prevalence $\rho^I$.  The larger $\gamma$, the more infected individuals turning back to the aware susceptible state, and hence the lower infection risk in the population. What is more, it follows from Eq.~\ref{fig4}(d) and Fig.~\ref{fig5}(d) that enhancing the adaptive rewiring mechanism for informed susceptible individuals would help prevent the disease outbreak by raising the disease threshold and mitigate the epidemic transmission by diminishing the disease prevalence $\rho^I$. This can be expounded as follows. According to the adaptive rewiring mechanism, parts of susceptible individuals are separated from infected ones. The greater the rewiring rate is, the more pathway for disease transmission will be interrupted, therefore the higher disease transmission threshold and the lower epidemic prevalence in case the disease prevails.

\begin{figure}[ht]
\begin{tabular}{cc}
  \mbox{\includegraphics*[width=0.45\columnwidth,clip=0]{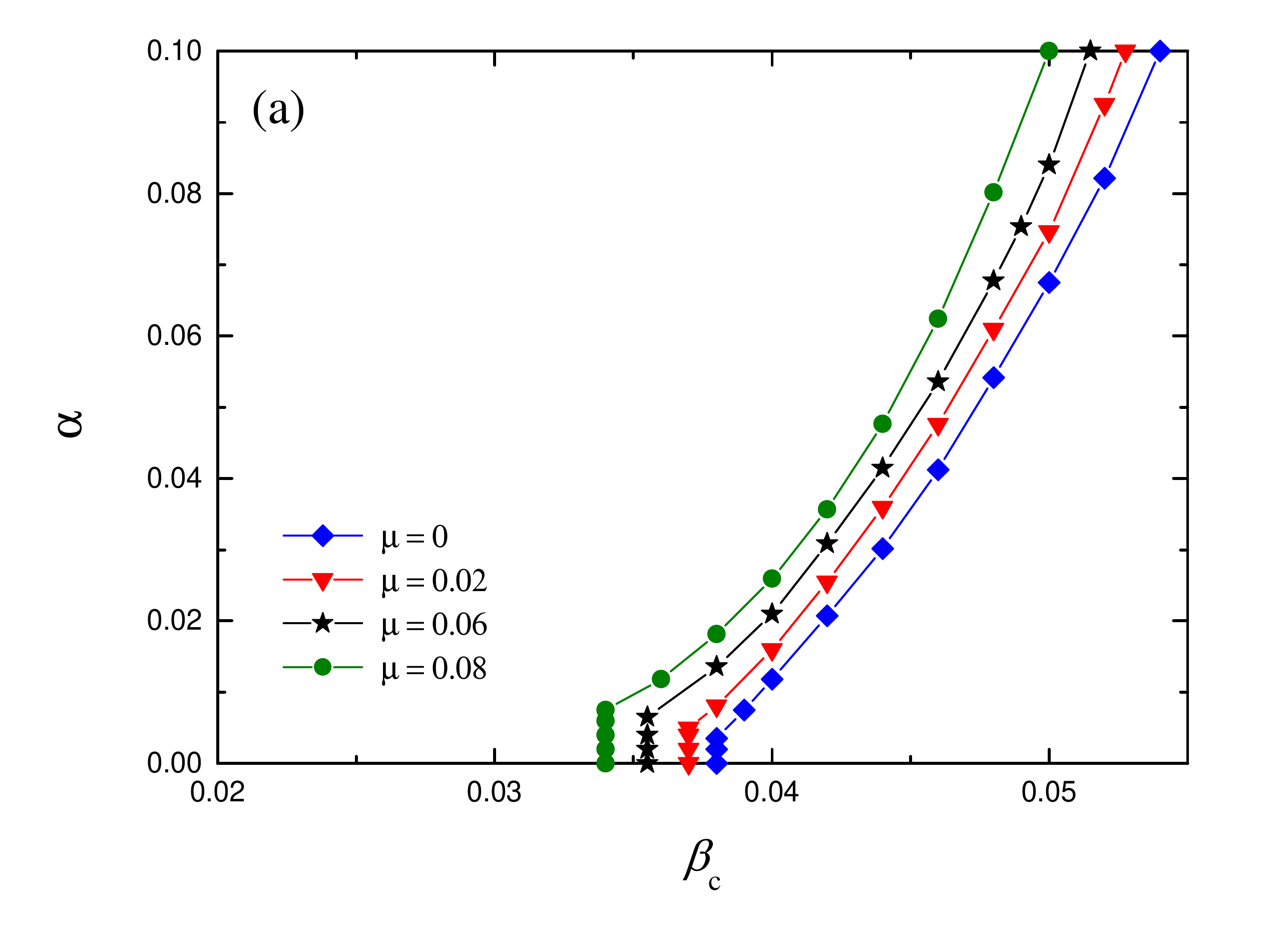}} &
  \mbox{\includegraphics*[width=0.45\columnwidth,clip=0]{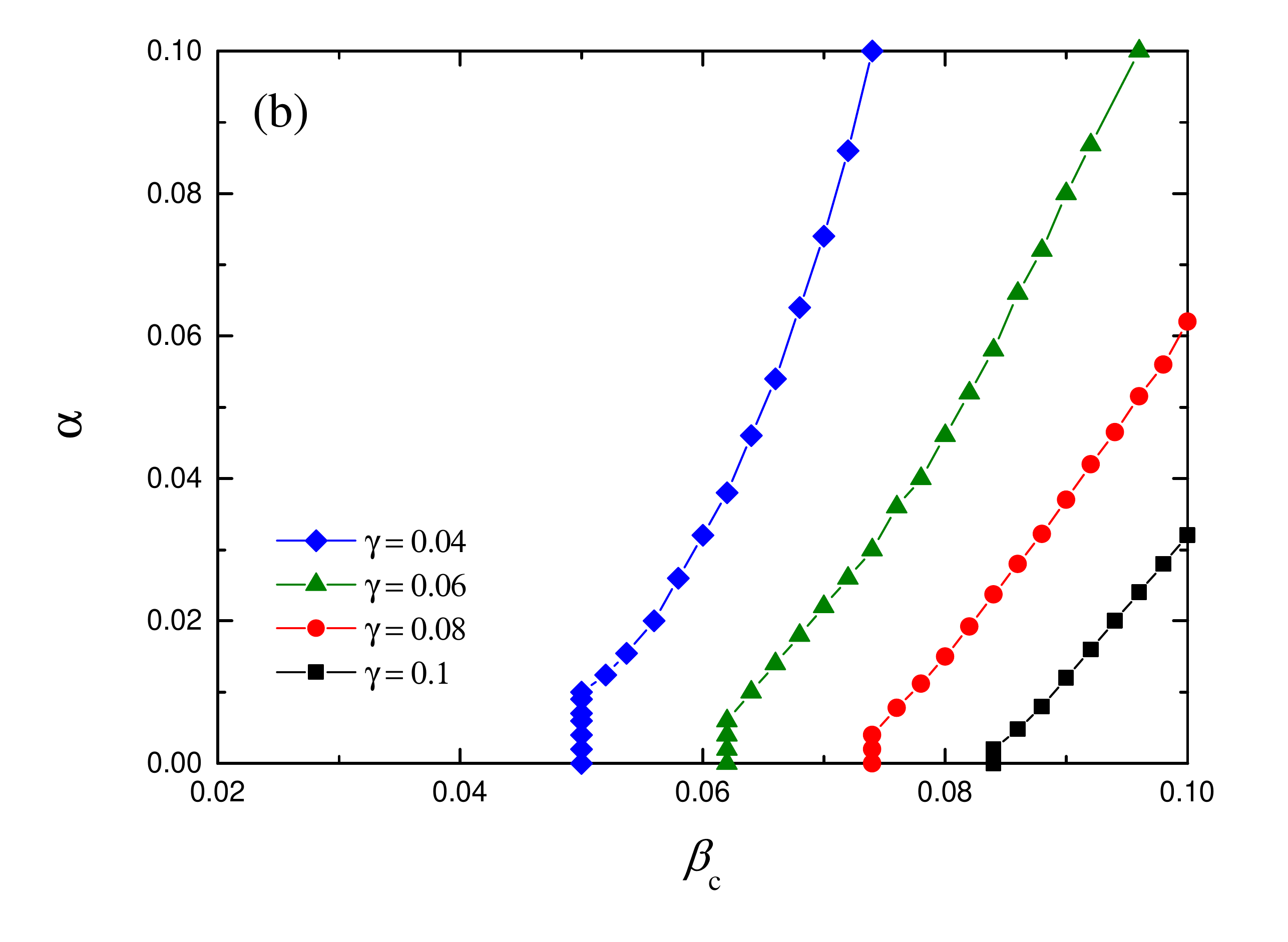}}
  \\
  \mbox{\includegraphics*[width=0.45\columnwidth,clip=0]{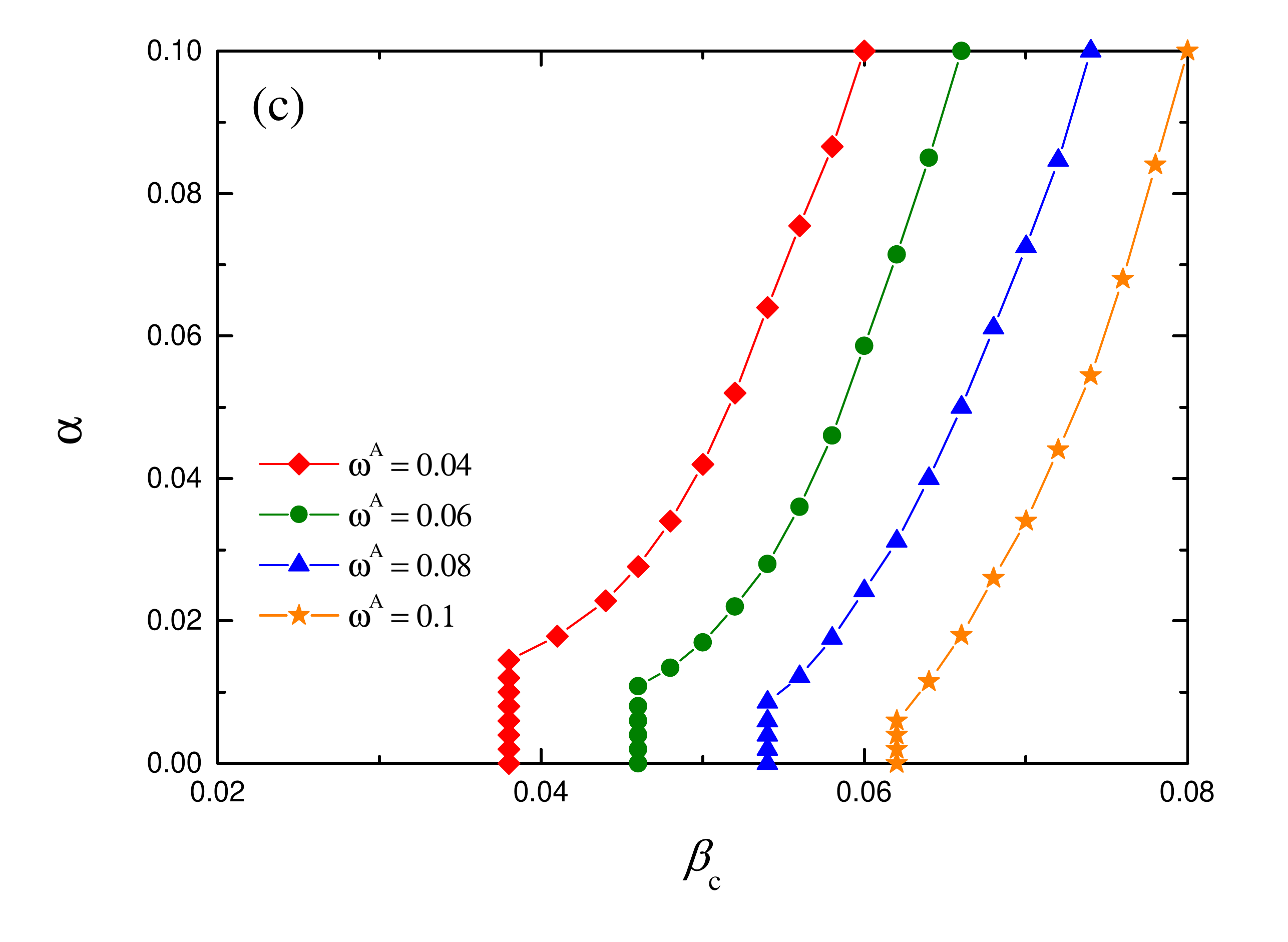}} &
  \mbox{\includegraphics*[width=0.45\columnwidth,clip=0]{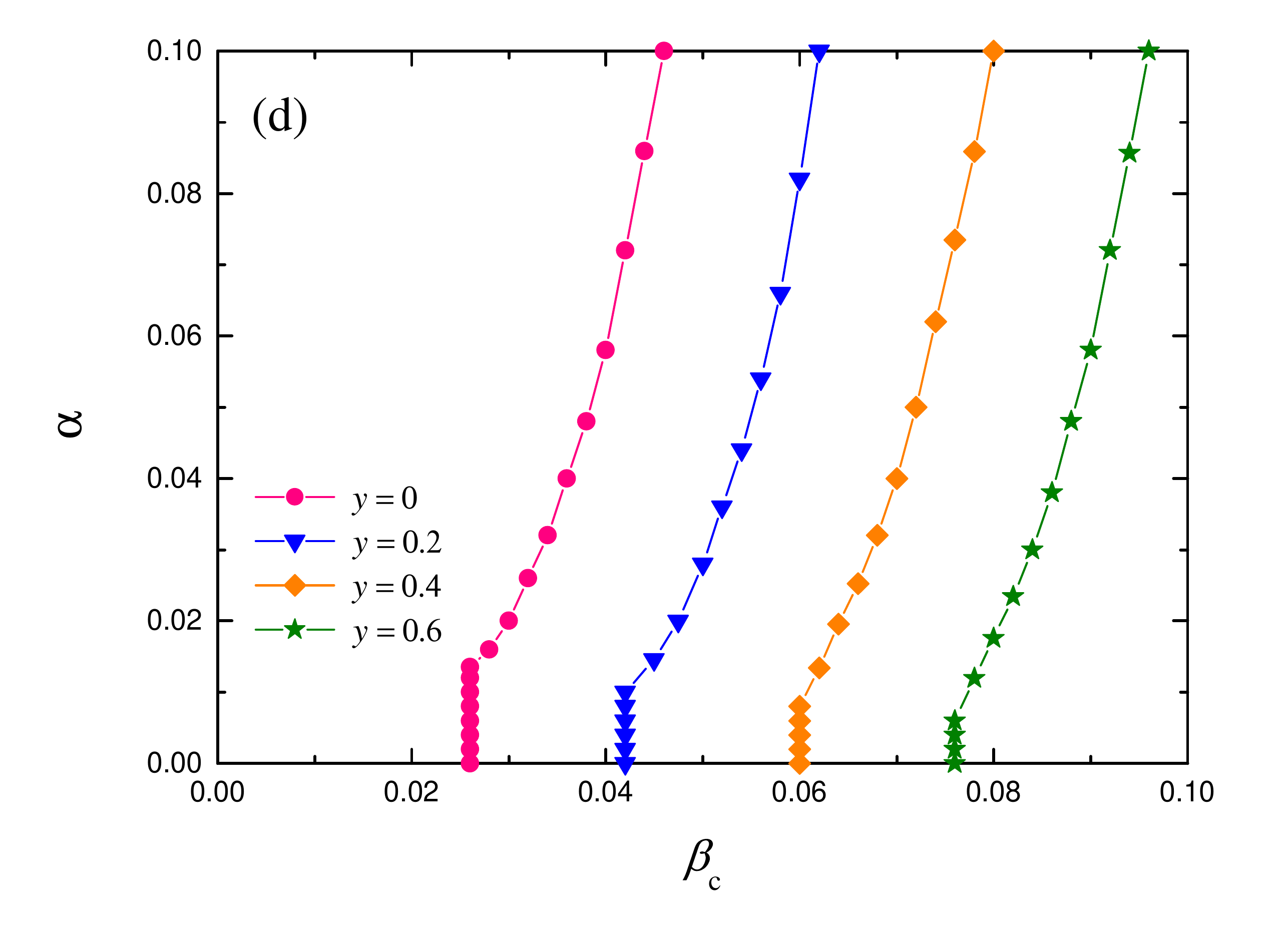}}
\end{tabular}
\caption{(color online) Critical transmission rate $\beta_{\rm c}$ for disease outbreak as a function of the awareness information spreading rate $\alpha$ with different values of (a) the information forgetting rate $\mu$, (b) the recovery rate $\gamma$ from infectious disease, (c) the rewiring rate $\omega^A$ for aware susceptible individuals, and (d) the ratio  $y=\omega^A/\omega^U$. All the results are obtained based on the basic reproduction number for disease transmission [Eq.~(\ref{eq16})]. The initial network setting is the same as in Fig.~\ref{fig5}. Other parameters are set for panels (a) $\gamma=0.02$, $\omega^A=0.05$, $\omega^U=0.01$; (b) $\mu=0.04$, $\omega^A=0.05$, $\omega^U=0.01$; (c) $\mu=0.06$, $\gamma=0.03$, $y=\omega^U/\omega^A=0.2$; (d) $\mu=0.06$, $\gamma=0.03$, $\omega^A=0.05$.}\label{fig7}
\end{figure}

In Fig.~\ref{fig6}, we provide a clear comparison for the disease outbreak condition between theoretical analysis and stochastic simulations. In particular, as denoted by the color bar in the heat map, the density $\rho^I$ of infected nodes in the steady state is mapped onto a two-parameter phase plane. In each panel of Fig.~\ref{fig6}, the theoretical results of disease outbreak threshold are tripartite. The disease threshold derived from Eq.~(\ref{eq16}) in the effective degree model is marked by solid curves, the disease threshold obtained from Eq.(\ref{eq26}) in the reduced model is represented by empty circles, and the lower bound of disease threshold calculated based on Eq.~(\ref{eq27}) is indicated by dashed curves. As shown in Fig.~\ref{fig6}(a), in the case of relatively low rewiring rate ($\omega^A=0.01$), the information diffusion rate $\alpha$ does not have much impact on the density $\rho^I$ of infected nodes, which is greatly affected by the disease transmission rate $\beta$. By contrast, in the case of relatively large rewiring rates [$\omega^A=0.08$ in Fig.~\ref{fig4}(b) and $\omega^A=0.05$ in Fig.~\ref{fig5}(b)], the information diffusion rate $\alpha$ has substantial effects on $\rho^I$.  This suggests that the combination of enhanced information propagation with forceful adaptive rewiring would play an important role in disease prevention and control. More importantly, though the lower bound of disease threshold is independent of $\alpha$, both the disease transmission thresholds $\beta_{\rm c}$ derived from our effective degree model and the reduced model demonstrate a metacritical point ($\alpha_{\rm c}, \beta_{\rm c}$) [marked by the red filled circle in Fig.~\ref{fig6}(a)] as previously reported in \cite{Gran13}. As indicated by solid curves and empty circles in Fig.~\ref{fig6}(a), there is a critical phenomenon in the dependence of the disease outbreak threshold $\beta_{\rm c}$ on the information spreading rate $\alpha$. Namely, there exists a critical value $\alpha_{\rm c}$ for information spreading rate, above which  $\beta_{\rm c}$ nonlinearly grows with the increase of $\alpha$, whereas below which $\beta_{\rm c}$ does not vary with $\alpha$. In Fig.~\ref{fig6}(b), the value of $\rho^I$ is plotted in the $\beta-\gamma$ phase plane, where the disease-free region (marked in blue) accounts for a major part, which highlights the great contribution of enlarging the recovery rate $\gamma$ to preventing the disease outbreak. The solid curve indicates that the disease transmission threshold $\beta_{\rm c}$ obtained from effective degree model [Eq.~(\ref{eq16})] grows nonlinearly with $\gamma$, whilst the disease transmission thresholds derived from the reduced model [Eq.~(\ref{eq26})] (empty circles) and the lower bound [Eq.~(\ref{eq27})] (dashed curves) increase almost linearly with $\gamma$. Only in the case of an extremely small $\gamma$ with a relatively large $\beta$ can the value $\rho^I$ reach the maximum. In Fig.~\ref{fig6}(c), it is shown that for the given parameters the disease transmission rate $\beta$ exhibits a dominant influence on the density $\rho^I$ of infected nodes over the information forgetting rate $\mu$, especially for large values of $\mu$. In addition, the disease outbreak threshold shows a negligible reliance on $\mu$ , which has also been demonstrated in Fig.~\ref{fig4}(a) and Fig.~\ref{fig5}(a). Furthermore, it is shown in Fig.~\ref{fig6}(d) that all the three theoretical disease thresholds grows almost linearly with the adaptive rewiring rate $\omega^A$. The higher the disease transmission rate, the more frequent adaptive rewiring needed for disease extinction. In general, as shown in all panels in Fig.~\ref{fig6}, the theoretical result of disease outbreak threshold matches well with the stochastic simulations. What is more, as expected, the threshold transmission rate $\beta_{\rm c}$ obtained by effective degree model is smaller than the one obtained in the reduced model, which is in turn less than that derived from the lower bound $\mathcal{R}_0^{\rm LB}$. It is worth noting that $\mathcal{R}_0^{\rm LB}>1$ ensures $\mathcal{R}_0^{\rm dis}>1$. Therefore, the lower bound can be used as an early warning signal of epidemic risk as well as a preliminary guidance for potential intervention measures for disease prevention.

\begin{figure}
\begin{center}
\includegraphics[width=0.9\columnwidth]{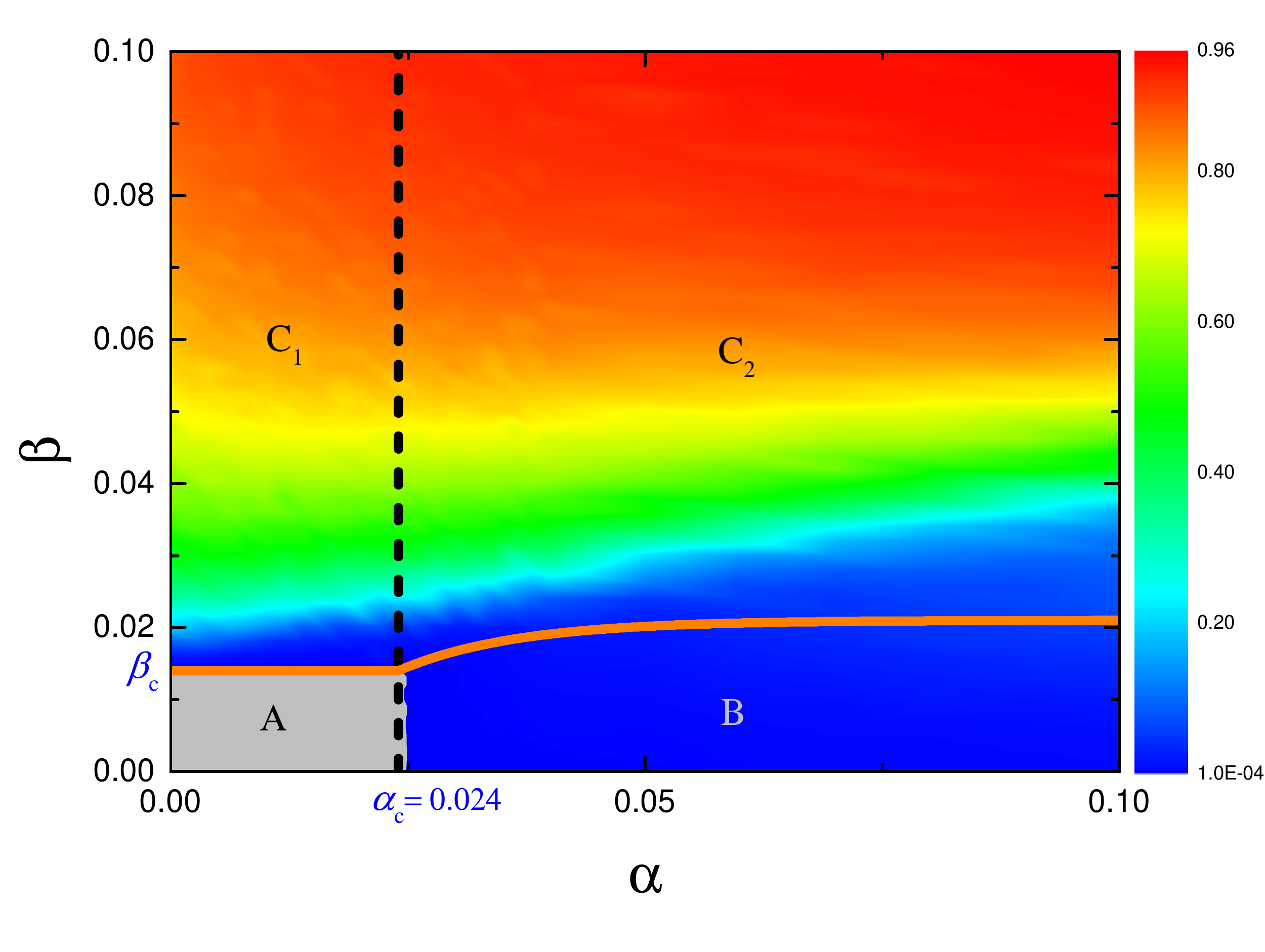}
\caption{(color online) Heat map for the density $\rho^A$ (denoted by the color bar) of informed individuals (including infected individuals and aware susceptible individuals) who are aware of disease information in the stationary state as plotted in the $\alpha-\beta$ phase diagram. The orange solid curve is for the disease outbreak threshold ($\beta=\beta_{\rm c}$) derived from Eq.~(\ref{eq16}), while the black dashed curve corresponds to the threshold condition ($\alpha=\alpha_{\rm c}$) for awareness information outbreak obtained based on Eq.~(\ref{eq22}). The $\alpha-\beta$ phase plane is divided by the threshold curves into four regions: the grey region A denotes the disease- and information-free steady state (i.e., there are only unaware susceptible individuals in the end); the blue region B denotes the disease-free but information-present steady state (i.e., both aware and unaware susceptible individuals exist without disease in the end); and the polychrome regions ${\rm C}_1$ and ${\rm C}_2$ mark the disease endemic state where infected individuals prevail in the population. The initial multiplex network structure is the same as in Fig.~\ref{fig5}. The simulation results are obtained by averaging over $100$ independent realizations. Other parameters are $\mu=0.02$, $\gamma=0.01$, $\omega^A=0.05$, $\omega^U=0.001$.}\label{fig8}
\end{center}
\end{figure}

In Fig.~\ref{fig7}, we present the disease transmission threshold $\beta_{\rm c}$ as a function of the the awareness information diffusion rate $\alpha$ with different values of other parameters. In all panes of Fig.~\ref{fig7}, there exhibits a metacritical point $(\alpha_{\rm c}, \beta_{\rm c})$ such that the value of $\beta_{\rm c}$ grows nonlinearly with $\alpha$ when $\alpha>\alpha_{\rm c}$; otherwise, $\beta_{\rm c}$ is independent of $\alpha$. This critical phenomenon in the reliance of disease transmission threshold on the awareness information diffusion rate coincides with the findings in \cite{Gran13,Kiss10,Huang18}. As shown in Fig.~\ref{fig7}(a), with parameters $\gamma=0.02$, $\omega^A=0.05$ and $\omega^U=0.01$, the disease threshold $\beta_{\rm c}$ decreases slightly as $\mu$ rises, which is consistent with the results in Figs.~\ref{fig4}(a), \ref{fig5}(a) and \ref{fig6}(c). Moreover, for any fixed $\alpha$, as displayed in Fig.~\ref{fig7}(a), the larger the information forgetting rate $\mu$, the smaller the disease threshold $\beta_{\rm c}$. Reversely, it is shown from Eq.~\ref{fig7}(b-d) that the disease transmission threshold $\beta_{\rm c}$ grows significantly with the rise of parameters $\gamma$, $\omega^A$ and $y$. That is, for any fixed information propagating rate $\alpha$, the larger the disease recovery rate and adaptive rewiring rates, the bigger the critical transmission rate for the onset of epidemic. In addition, in combination with a large $\alpha$, increasing the rates of recovery and rewiring leads to a greater $\beta_{\rm c}$, which highlights the combined effects of awareness information spreading and adaptive rewiring in preventing and suppressing the disease transmission. Furthermore, it is worth noting that the critical value $\alpha_{\rm c}$ of awareness spreading rate (for the increase of the epidemic onset $\beta_{\rm c}$) grows accordingly with the increase of $\mu$ and with the decrease of $\gamma$, $\omega^A$ and $y$. In particular, as shown in each panel of Fig.~\ref{fig7}, the smaller $\beta_{\rm c}$, the larger $\alpha_{\rm c}$.

According to our theoretical analysis on the threshold conditions for disease outbreak and awareness information invasion in \ref{sec4}, it follows that if $\beta>\beta_{\rm c}$ (which is equivalent to $\mathcal{R}_0^{\rm dis}>1$) then the disease will break out; otherwise, the disease will die out. On the other hand, if $\alpha>\alpha_{\rm c}$ (which is equivalent to $\mathcal{R}_0^{\rm awe}>1$) then the awareness information will prevail in the population; otherwise, the awareness information can not invade through the information layer. Besides, based on the stability analysis by Kiss \textit{et al.} \cite{Kiss10} and Huang \textit{et al.} \cite{Huang18} on the similar awareness-disease model, it is claimed that when $\mathcal{R}_0^{\rm dis}\leq 1$ and $\mathcal{R}_0^{\rm awe}\leq 1$, the disease- and information-free equilibrium is stable; when $\mathcal{R}_0^{\rm dis}\leq 1$ and $\mathcal{R}^{\rm awe}_0>1$, the disease-free and information statured equilibrium is stable; in the case of $\mathcal{R}_0^{\rm dis}>1$, whether the disease eventually dies out or not depends on the competition between the two processes of awareness information diffusion and disease transmission. In particular, if $\mathcal{R}^{\rm dis}_0>1$ and $\mathcal{R}_0^{\rm awe}\leq 1$, then the disease spreading outweighs the information diffusion process, the disease prevails; otherwise, if the information diffusion outweighs the disease spreading, then awareness information prevails.  Note that if the disease prevails in the population then, based on our model definition, the infected individuals will immediately become aware of the disease information. Therefore, in the case of $\mathcal{R}_0^{\rm dis}>0$ the awareness information prevails in the population.

In order to verify the above statement, we present in Fig.~\ref{fig8} the heat map for the density $\rho^A$ of aware individuals who know of the disease information as a function of spreading rates $\alpha$ of awareness information and $\beta$ of disease. Note that the grey region A marks the information-free state (i.e., $\rho^A=0$) and the $\rho^A$ value goes from $0.0001$ to $0.96$ as the color changes gradually from blue to red, as denoted in the color bar. For comparison, the critical conditions for the onset of disease and awareness information have been depicted in the $\alpha-\beta$ plane. The orange solid curve represents the disease threshold $\beta_{\rm c}$ below which the disease will die out, whereas the black dashed curve stands for the awareness information threshold $\alpha_{\rm c}$ above which the awareness information prevails in the population. Consequently, the $\alpha-\beta$ plane is divided by the threshold curves into four areas A, B, ${\rm C}_1$ and ${\rm C}_2$. The simulation results agree well with our theoretical analysis in such a way that in area A, there is no disease and no awareness information at all; in area B, the disease dies out but the information persists in the population; in both areas ${\rm C}_1$ and ${\rm C}_2$, the awareness information prevails with a relative higher level, albeit whether the disease dies out or not in area ${\rm C}_2$ depends on the outcome of the competition between disease transmission and awareness information diffusion \cite{Kiss10,Huang18}. It can be observed from Fig.~\ref{fig8} that the critical value of information diffusion rate in the metacritical point mentioned in Fig.~\ref{fig6}(a) and Fig.~\ref{fig7} is exactly the awareness information threshold $\alpha_{\rm c}$, above which the onset of disease grows.

\section{Conclusions}\label{sec7}

Summarizing, in this paper we have investigated the effects of awareness propagation and individuals' contact behavioral changes on infectious disease transmission over multiplex networks. The risk-avoiding contact changes are characterized by adaptive link rewiring that is supposed to be dependent on individuals' awareness state about the disease. To account for the structural evolution of the contact network caused by the adaptive link rewiring process, we have developed an effective-degree contagion dynamics model incorporating an SIS epidemic spreading process and a UAU awareness diffusion process. Employing the next generation matrix approach \cite{vW02}, we derived the disease reproduction number $\mathcal{R}_0^{\rm dis}$ and awareness information reproduction number $\mathcal{R}_0^{\rm awe}$ for our effective degree model, based on which we obtained the threshold rates $\beta_{\rm c}$ for disease transmission and $\alpha_{\rm c}$ for awareness information diffusion. In addition, through a reduced model, two lower bounds of the basic reproduction number have been provided, which could be used as a warning signal to assess the risk of infections and to provide early preparedness for disease prevention. Moreover, the effects of various parameters on the disease threshold as well as the epidemic prevalence have been explored via extensive simulations. The critical phenomenon is observed in the dependence of the onset of epidemics on the awareness information diffusion rate, which supports the emergence of a metacritical point
reported by Granell \textit{et al.} \cite{Gran13}. Furthermore, it is shown from our results that the awareness-dependent link rewiring contributes to enlarging the disease transmission threshold $\beta_{\rm c}$ and lowering the epidemic prevalence, which highlights the effects of adaptive rewiring on preventing the disease spread. Besides, the results also suggest that the combination of intensified awareness diffusion and enhanced link rewiring will lead to greater contribution for disease prevention and control. Finally, we have also presented a two-parameter bifurcation diagram for the density of aware individuals in the stationary state, providing a numerical support for the stability analysis in the SIS-UAU contagion models \cite{Kiss10,Huang18}. It is to be noted that although the proposed effective-degree model allows for detailed description of changes in the connectivity structure, its drawback mainly lies in the computational complexity due to a complicated high-dimensional dynamical systems. Therefore, it is not appropriate to apply this model to a complex network with a number of high-degree nodes. It has long been challenging to strike the right balance between fine characterizations of dynamical processes with feasible computation complexities.

\addcontentsline{toc}{chapter}{Appendix A: Appendix section heading}
\section*{Appendix A: Proof of inequality $\mathcal{R}_0^{\rm dis} < \widetilde{\mathcal{R}}_0^{\rm dis}$}
\setcounter{equation}{0}
\renewcommand{\theequation}{A\arabic{equation}}

To proof $\mathcal{R}_0^{\rm dis} < \widetilde{\mathcal{R}}_0^{\rm dis}$, it is equivalent to show
\begin{equation}\label{eq29}
\frac{\widetilde{f}_2\widetilde{f}_3 }{\widetilde{f}_1\widetilde{f}_2\widetilde{f}_3-\gamma\beta(\mu+\widetilde{f}_1)}
> \frac{ f_2f_3 }{f_1f_2f_3-\gamma\beta(\mu+f_1)}.
\end{equation}
Since the denominators on both sides of the above inequality ({\ref{eq29}}) are positive,  what we need is to show
\begin{equation}
\Delta_{k}\triangleq f_1 f_2\Big[{\widetilde{f}_1\widetilde{f}_2\widetilde{f}_3-\gamma\beta(\mu+\widetilde{f}_1)}\Big]  - {\widetilde{f}_2\widetilde{f}_3 }\Big[{f_1f_2f_3-\gamma\beta(\mu+f_1)}\Big] < 0.
\end{equation}
Substituting the expressions of $\widetilde{f}_1$, $\widetilde{f}_2$ and $\widetilde{f}_3$ into $\Delta_{k}$, then expanding all the terms, and letting $\Lambda=k-1\geq 0$ gives rise to
\begin{widetext}
\begin{eqnarray}
\Delta_{k}&=&-\Lambda^2\alpha^2\beta^2n^2\omega^{U}-\Lambda^2\alpha^2\beta n^2(\omega^{U})^2-2\Lambda\alpha^3\beta{n^3}\omega^{U}
-\Lambda\alpha^3{n^3}(\omega^{U})^2-\alpha^4{n^4}\omega^{U}-2\Lambda^2\alpha\beta^3n\omega^{U}-2\Lambda^2\alpha\beta^2\gamma n\omega^{U} \nonumber \\
&& -2\Lambda^2\alpha\beta^2\mu n\omega^{U}-\Lambda^2\alpha\beta^2 n\omega^{A}\omega^{U}-2\Lambda^2\alpha\beta^2n(\omega^U)^2-2\Lambda^2\alpha\beta\gamma n(\omega^U)^2-2\Lambda^2\alpha\beta\mu n(\omega^U)^2-\Lambda^2\alpha\beta n\omega^A(\omega^U)^2 \nonumber\\
&& -4\Lambda\alpha^2\beta^2 n^2 \omega^U-8\Lambda\alpha^2\beta\gamma n^2 \omega^{U} -4\Lambda \alpha^2\beta\mu n^2 \omega^{U}
-2\Lambda\alpha^2\beta n^2 \omega^A\omega^U-2\Lambda\alpha^2\beta n^2 (\omega^U)^2 -4\Lambda\alpha^2\gamma n^2(\omega^U)^2 \nonumber\\
&& -2\Lambda\alpha^2\mu n^2(\omega^U)^2 - \Lambda\alpha^2 n^2\omega^{A}(\omega^U)^2 -2\alpha^3\beta n^3 \omega^U -6\alpha^3\gamma n^3 \omega^U -2\alpha^3\mu n^3 \omega^U - \alpha^3 n^3 \omega^A\omega^U - \Lambda^2\beta^4\omega^U \nonumber\\
&& -2\Lambda^2\beta^3\gamma\omega^U - 2\Lambda^2\beta^3\mu\omega^U - \Lambda^2\beta^3\omega^A\omega^U - \Lambda^2\beta^3(\omega^U)^2 -\Lambda^2\beta^2\gamma^2\omega^U - 2\Lambda^2\beta^2\gamma\mu\omega^U -\Lambda^2\beta^2\gamma\omega^A\omega^U \nonumber\\
&& -2\Lambda^2\beta^2\gamma(\omega^U)^2 -\Lambda^2\beta^2\mu^2\omega^U -\Lambda^2\beta^2\mu\omega^A\omega^U - 2\Lambda^2\beta^2\mu(\omega^U)^2 -\Lambda^2\beta^2\omega^A(\omega^U)^2 -\Lambda^2\beta\gamma^2(\omega^U)^2 \nonumber\\
&& -2\Lambda^2\beta\gamma\mu(\omega^U)^2 -\Lambda^2\beta\gamma\omega^A(\omega^U)^2 - \Lambda^2\beta\mu^2(\omega^U)^2 - \Lambda^2\beta\mu\omega^A(\omega^U)^2 - 2\Lambda\alpha\beta^3 n\omega^U -12\Lambda\alpha\beta^2 \gamma n\omega^U \nonumber\\
&& -4\Lambda \alpha\beta^2\mu n \omega^U -2\Lambda\alpha\beta^2 n\omega^A\omega^U -\Lambda\alpha\beta^2 n(\omega^U)^2 -11\Lambda \alpha\beta\gamma^2 n\omega^U -13\Lambda\alpha\beta\gamma\mu n\omega^U -6\Lambda\alpha\beta\gamma n\omega^A\omega^U \nonumber\\
&& -6\Lambda\alpha\beta\gamma n(\omega^U)^2 - 2\Lambda\alpha\beta\mu^2 n\omega^U - 2\Lambda\alpha\beta\mu n\omega^A\omega^U -2\Lambda\alpha\beta\mu n(\omega^U)^2 -\Lambda\alpha\beta n\omega^A(\omega^U)^2 -5\Lambda\alpha\gamma^2 n(\omega^U)^2 \nonumber\\
&& -6\Lambda\alpha\gamma\mu n(\omega^U)^2 -3\Lambda\alpha\gamma n\omega^A(\omega^U)^2 -\Lambda\alpha\mu^2 n(\omega^U)^2 -\Lambda\alpha\mu n\omega^A(\omega^U)^2 -\alpha^2\beta^2n^2\omega^U -9\alpha^2\beta\gamma n^2\omega^U \nonumber\\
&& -2\alpha^2\beta\mu n^2\omega^U -\alpha^2\beta n^2\omega^A\omega^U -13\alpha^2\gamma^2 n^2\omega^U -10\alpha^2\gamma\mu n^2\omega^U -5\alpha^2\gamma n^2 \omega^A\omega^U -\alpha^2\mu^2 n^2\omega^U -\alpha^2\mu n^2\omega^A\omega^U \nonumber\\
&& -\Lambda\beta^3\gamma\omega^A -4\Lambda\beta^3\gamma\omega^U -\Lambda\beta^2\gamma^2\omega^A -9\Lambda\beta^2\gamma^2\omega^U -\Lambda\beta^2\gamma\mu\omega^A -9\Lambda\beta^2\gamma\mu\omega^U -5\Lambda\beta^2\gamma\omega^A\omega^U -2\Lambda\beta^2\gamma(\omega^U)^2 \nonumber \\
&& -5\Lambda\beta\gamma^3\omega^U -10\Lambda\beta\gamma^2\mu\omega^U -5\Lambda\beta\gamma^2\omega^A\omega^U -4\Lambda\beta\gamma^2(\omega^U)^2 -5\Lambda\beta\gamma\mu^2\omega^U -5\Lambda\beta\gamma\mu\omega^A\omega^U
-4\Lambda\beta\gamma\mu(\omega^U)^2 \nonumber\\
&& -2\Lambda\beta\gamma\omega^A(\omega^U)^2 -2\Lambda\gamma^3(\omega^U)^2 -4\Lambda\gamma^2\mu(\omega^U)^2 -2\Lambda\gamma^2\omega^A(\omega^U)^2 -2\Lambda\gamma\mu^2(\omega^U)^2 -2\Lambda\gamma\mu\omega^A(\omega^U)^2 \nonumber\\
&& -\alpha\beta^2\gamma n\omega^A -3\alpha\beta^2\gamma n\omega^U -\alpha\beta\gamma^2 n\omega^A -13\alpha\beta\gamma^2 n\omega^U -\alpha\beta\gamma\mu n\omega^A -7\alpha\beta\gamma\mu n\omega^U -4\alpha\beta\gamma n\omega^A\omega^U \nonumber\\
&& -12\alpha\gamma^3 n\omega^U -16\alpha\gamma^2\mu n\omega^U -8\alpha\gamma^2 n\omega^A\omega^U -4\alpha\gamma\mu^2 n\omega^U -4\alpha\gamma\mu n\omega^A\omega^U -2\beta^2\gamma^2\omega^A -2\beta^2\gamma^2\omega^U \nonumber\\
&& -2\beta\gamma^3\omega^A -6\beta\gamma^3\omega^U -2\beta\gamma^2\mu\omega^A -6\beta\gamma^2\mu\omega^U -4\beta\gamma^2\omega^A\omega^U -4\gamma^4\omega^U -8\gamma^3\mu\omega^U -4\gamma^3\omega^A\omega^U \nonumber\\
&& -4\gamma^2\mu^2\omega^U -4\gamma^2\mu\omega^A\omega^U.
\end{eqnarray}
\end{widetext}
Evidently,  $\Delta_{k}<0$ holds for any $k\geq 1$. The proof of the inequality $\mathcal{R}_0^{\rm dis} < \widetilde{\mathcal{R}}_0^{\rm dis}$ is complete.

\addcontentsline{toc}{chapter}{Appendix A: Appendix section heading}
\section*{Appendix B: Impact of rewiring on degree distribution of the epidemic-layer network}
\begin{figure}
\begin{center}
\includegraphics[width=0.9\columnwidth]{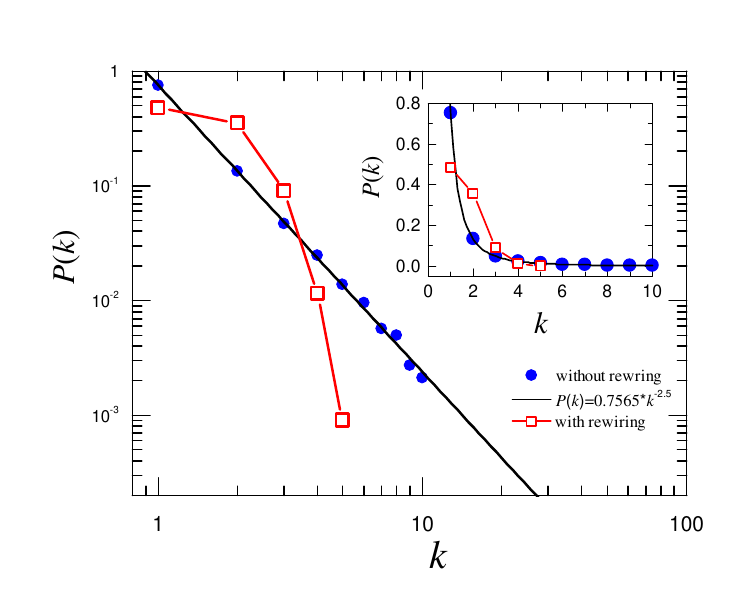}
\caption{(color online) Degree distributions of the underlying network capable of disease transmission in the case without (blue solid circles) and with (red empty squares) adaptive rewiring, respectively, at time $t=10^4$ when the system is in the stationary state. Black solid line denotes the exact power law with exponent $\nu=2.5$. The simulation results (denoted by symbols) are obtained by starting with the same initial network architecture as in Fig.~\ref{fig3}. To make a direct comparison between the cases without and with adaptive rewiring, the results are plotted in the log-log scale, while the inset displays the same results in a linear-linear scale. Parameters are $\alpha=0.04$, $\mu=0.1$, $\gamma=0.02$, $\beta=0.2$, $\omega^A=0.02$, $\omega^U=0.004$.}\label{fig9}
\end{center}
\end{figure}

According to the adaptive rewiring rule in our model, the underlying network structure in the epidemic layer---especially the degree distribution---will evolve along with the rewiring process. Adaptive rewiring of high-risk links leads to the breakdown of edges that connect susceptible nodes and infected ones, and meanwhile, gives rise to the formation of low-risk links connecting towards a randomly chosen susceptible node. The random attaching mechanism that has been widely used in random graph models \cite{ER59} will typically generate a relatively narrow degree distribution. As demonstrated in Fig.~\ref{fig9} the degree distribution of underlying contact network in the epidemic layer exhibits different scaling behaviors: in the case without rewiring, the node degree follows a perfect power law with exponent $\nu=2.5$; while in the case with rewiring, the degree values are narrowly distributed along the average degree, approximating to a Poisson distribution.





\end{document}